\documentclass[fleqn,usenatbib]{mnras}

\usepackage[T1]{fontenc}

\DeclareRobustCommand{\VAN}[3]{#2}
\let\VANthebibliography\thebibliography
\def\thebibliography{\DeclareRobustCommand{\VAN}[3]{##3}\VANthebibliography}

\usepackage{pdflscape}

\usepackage{graphicx}	
\usepackage{amsmath}	
\usepackage{amssymb}	
\usepackage{caption}
\usepackage{subcaption}
\usepackage{booktabs}
\usepackage{multirow}

\newcommand{\jks}{\mbox{$J\!-\!K_{\rm s}$}}
\newcommand{\yks}{\mbox{$Y\!-\!K_{\rm s}$}}

\newcommand{\ks}{\mbox{$K_{\rm s}$}}

\newcommand{\av}{\mbox{$A_V$}}

\newcommand{\feh}{\mbox{\rm [{\rm Fe}/{\rm H}]}}

\newcommand{\Msun}{\mbox{$\mathrm{M}_{\odot}$}}

\newcommand{\Teff}{\mbox{$T_{\rm eff}$}}

\newcommand{\logg}{\mbox{$\log g$}}
\newcommand{\logtyr}{\mbox{$\log(t/\mathrm{yr})$}}
\newcommand{\SFRt}{\mbox{$\mathrm{SFR}(t)$}}

\newcommand{\lnL}{\mbox{$\ln \mathcal{L}$}}

\defcitealias{pastorelli19}{Paper~I}
\defcitealias{HZ09}{HZ09}

\title[SFH of LMC from VMC]{The VMC survey -- XLIII. The spatially resolved star formation history across the Large Magellanic Cloud}

\author[Mazzi et al.]{Alessandro Mazzi$^{1}$\thanks{E-mail:  alessandro.mazzi.phd@gmail.com},
L\'eo Girardi$^{2}$,
Simone Zaggia$^{2}$,
Giada Pastorelli$^{3}$, \newauthor 
Stefano Rubele$^{2}$, 
Alessandro Bressan$^{4}$,
Maria-Rosa L. Cioni$^{5}$, 
Gisella Clementini$^{6}$, \newauthor
Felice Cusano$^{6}$, 
Jo\~ao Pedro Rocha$^{7}$, 
Marco Gullieuszik$^{2}$, 
Leandro Kerber$^{7}$, \newauthor
Paola Marigo$^{1}$, 
Vincenzo Ripepi$^{8}$,
Kenji Bekki$^{9}$,
Cameron P.M. Bell$^{5}$, \newauthor
Richard de Grijs$^{10,11}$, 
Martin A. T. Groenewegen$^{12}$, 
Valentin D. Ivanov$^{13,14}$, \newauthor
Joana M. Oliveira$^{15}$, 
Ning-Chen Sun$^{16}$, 
Jacco Th. van Loon$^{15}$
\\
$^{1}$Dipartimento di Fisica e Astronomia Galileo Galilei, Universit\`a di Padova, Vicolo dell'Osservatorio 3, I-35122 Padova, Italy \\
$^{2}$INAF -- Osservatorio Astronomico di Padova, Vicolo dell'Osservatorio 5, I-35122 Padova, Italy \\
$^{3}$STScI, 3700 San Martin Drive, Baltimore, MD 21218, USA \\
$^{4}$SISSA, via Bonomea 365, I-34136 Trieste, Italy\\
$^{5}$Leibniz-Institut f\"ur Astrophysik Potsdam, An der Sternwarte 16, D-14482 Potsdam, Germany \\
$^{6}$INAF -- Osservatorio di Astrofisica e Scienza dello Spazio, Via Gobetti 93/3, I-40129 Bologna, Italy \\
$^{7}$Universidade Estadual de Santa Cruz, Depto.\ de Ci\^encias Exatas e Tecnol\'ogicas
Rodovia Jorge Amado km 16, 45662-900, Ilh\'eus, Brazil \\
$^{8}$INAF -- Osservatorio Astronomico di Capodimonte, Salita Moiariello 16, 80131, Naples, Italy \\
$^{9}$ICRAR, M468, The University of Western Australia, 35 Stirling Highway, Crawley, WA 6009, Australia \\
$^{10}$Department of Physics and Astronomy, Macquarie University, Balaclava Road, Sydney, NSW 2109, Australia \\
$^{11}$Research Centre for Astronomy, Astrophysics and Astrophotonics, Macquarie University, Balaclava Road, Sydney, NSW 2109, Australia \\
$^{12}$Koninklijke Sterrenwacht van Belgi\"e, Ringlaan 3, 1180 Brussels, Belgium \\
$^{13}$European Southern Observatory, Ave. Alonso de C\'ordova 3107, Vitacura, Santiago, Chile \\
$^{14}$European Southern Observatory, Karl-Schwarzschild-Str.2, 85748 Garching bei M\"unchen, Germany \\
$^{15}$Lennard-Jones Laboratories, Keele University, ST5 5BG, UK \\
$^{16}$Department of Physics and Astronomy, University of Sheffield, Hicks Building, Hounsfield Road, Sheffield S3 7RH, UK
}

\date{Accepted XXX. Received YYY; in original form ZZZ}

\pubyear{2021}

\usepackage{newtxtext,newtxmath}

\begin{document}
\label{firstpage}
\pagerange{\pageref{firstpage}--\pageref{lastpage}}
\maketitle

\begin{abstract}
We derive the spatially-resolved star formation history (SFH) for a $96$~deg$^2$ area across the main body of the Large Magellanic Cloud (LMC), using the near-infrared photometry from the VISTA survey of the Magellanic Clouds (VMC). The data and analyses are characterised by a great degree of homogeneity and a low sensitivity to the interstellar extinction. 756 subregions of size $0.125$~deg$^2$ --  corresponding to projected sizes of about $296\times322\,\mathrm{pc}^{2}$ in the LMC -- are analysed.
The resulting SFH maps, with typical resolution of $0.2$--$0.3$~dex in logarithm of age, reveal main features in the LMC disc at different ages: the patchy star formation at recent ages, the concentration of star formation on three spiral arms and on the Bar up to ages of $\sim\!1.6$~Gyr, and the wider and smoother distribution of older populations. The period of most intense star formation occurred roughly between 4 and 0.5 Gyr ago, at rates of $\sim\!0.3\,\Msun\mathrm{yr}^{-1}$. We compare young and old star formation rates with the observed numbers of RR Lyrae and Cepheids. We also derive a mean extinction and mean distance for every subregion, and the plane that best describes the spatial distribution of the mean distances. 
Our results cover an area about 50 per cent larger than the classical SFH maps derived from optical data by \citet{HZ09}. Main differences with respect to those maps are lower star formation rates at young ages, and a main peak of star formation being identified at ages slightly younger than $1$~Gyr. 
\end{abstract}




\section{Introduction}

Spatially resolved maps of the star formation history (SFH) of nearby galaxies are important essentially for two reasons: because they help to reconstruct the history of the Local Group, and because they help to improve the current theories of stellar evolution and stellar populations. And, among all nearby galaxies for which spatially resolved SFHs can be derived, the Large Magellanic Cloud (LMC) stands out as a primary target, given its large angular size, proximity ($\sim\!50$~kpc), and relatively simple geometry. 
Indeed, the external areas of the LMC can have their SFHs studied using the most simple data and methods available, namely optical ground-based photometry together with colour-magnitude diagram (CMD) reconstruction assuming a mix of stellar populations at a single-distance and low extinction. The inner and more crowded areas, instead, require sharper imaging -- which is progressively being expanded with better ground-based surveys and Hubble Space Telescope (HST) observations -- and a proper consideration of dust extinction during the SFH derivation.

Descriptions of the past history of the LMC (and the Magellanic System as a whole) include a series of key insights, such as 
an outside-in quenching of the star formation in fields between 2 and 6 kpc from the LMC centre \citep{gallart08,meschin14}, a reduced global star formation taking place until $5$--$3.5$~Gyr ago (\citealt{HZ09}, hereafter \citetalias{HZ09}; \citealt{weisz13}), the apparent coupling between field and cluster formation modes \citepalias{HZ09}, and that a common SFH is shared by the LMC Bar and by the inner LMC disc over timescales of gigayears \citep{monteagudo18}.
The addition of accurate proper motions \citep{kalli13,gaia_LMC} is creating substantial challenges for the interpretation of these SFHs -- but, on the other hand, it is opening the possibility of detailed comparisons between the SFHs of model galaxies derived from cosmological simulations \citep[e.g.][]{williamson21}, and those actually observed in the Magellanic Clouds.

Spatially resolved SFHs are also starting to become an important ingredient to test and improve the theories of stellar evolution and stellar populations. The simplest application of this kind stands on the delay-time distribution (DTD) technique, which relates the counts of a given class of objects with the SFHs of the galaxy region where they are observed. When many different galaxy regions are available, delay-times between the star formation events and the appearance of the objects can be derived, and inform about their lifetimes and progenitor masses. DTDs have so far been applied in the LMC to probe the progenitors of supernova remnants \citep{maoz10}, planetary nebulae \citep{badenes15}, and even RR Lyrae \citep{sarbad21}. Another technique relies on the fact that some classes of objects derive from a wide range of stellar masses, whose relative contributions can only be assessed if we have spatially resolved SFHs; comparisons between the numbers modelled and those observed in regions of different mean metallicity and different mean age, then inform us on the correctness of model lifetimes, and suggest directions for their improvement. This method requires that the SFH derivation is not affected by the stars whose models are being checked. Examples of this method applied to calibrate evolutionary models of TP-AGB stars in the Magellanic Clouds are given in \citet{pastorelli19,pastorelli20}.

The results of all these methods depend on the accuracy, reliability, and spatial extension of the derived SFHs. For the main body of the LMC, the classical space-resolved SFH map comes from \citetalias{HZ09}, and was derived from $\sim\!64$~deg$^2$ of optical photometry from the Magellanic Clouds Photometric Survey \citep[MCPS;][]{zaritsky04}. Different surveys are now aiming to improve these maps by using either deeper optical data, such as the Survey of the MAgellanic Stellar History \citep[SMASH;][]{smash1,smash2}, or the near-infrared data, such as the VISTA\footnote{VISTA is the  Visible and InfraRed Survey Telescope for Astronomy \citep{vista}, a 4-m telescope located at the Cerro Paranal site of the European Southern Observatory (ESO).} survey of the Magellanic Clouds \citep[VMC;][]{cioni11}.

Last but not least, the LMC is, traditionally, a main anchor in the measurement of the Hubble constant (H$_{0}$) through the cosmic distance ladder. A consistent and homogeneous SFH derived over the whole body of the LMC is crucial to disentangle how the LMC morphology impacts distances measured using most important and widely used standard candles such as Cepheids, RR Lyrae and eclipsing binaries.

In this paper, we derive the space-resolved SFH across the disc of the LMC using 63 tiles of VMC data for a total area of $96$~deg$^2$.
This work supersedes the preliminary results presented in \citet{rubele12} and \citet{pastorelli20} for smaller subsets of the VMC data for the LMC (regarding 4 and 11 tiles, respectively), and complements the analyses of the SFH for the Small Magellanic Cloud (SMC) using the same survey \citep{rubele15, rubele18}. A distinctive characteristic of the present analysis is the great uniformity of the entire dataset, which we match with a uniform method of analysis. In addition, the use of near-infrared data ensures a reduced sensitivity of the results on the extinction, both internal and external to the LMC.

This paper is structured as follows: Section~\ref{sec:obs} presents the VMC data and its processing for the aims of this paper. 
Section~\ref{sec:method} describes the method we adopt to derive the SFH together with examples for a couple of LMC subregions. 
Section~\ref{sec:sfh} presents the results for the large-scale map of the SFH, and the distances and extinctions we derive as a by-product of the method. 
Section~\ref{sec:discu} presents some additional analyses, namely a first order description of the LMC geometry based on the distance map and a comparison of the SFH map with the HZ09 one. 
Section~\ref{sec:conclu} summarises the main results.

\begin{figure*}
	\includegraphics[width=\columnwidth]{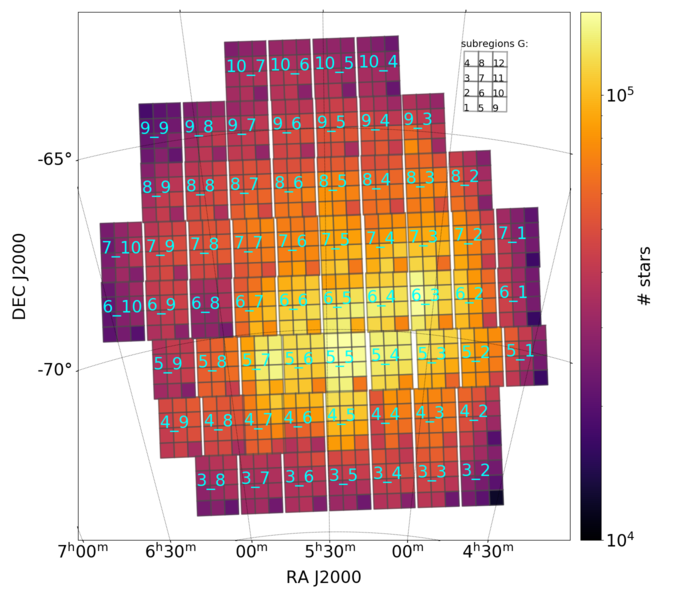}
	\includegraphics[width=\columnwidth]{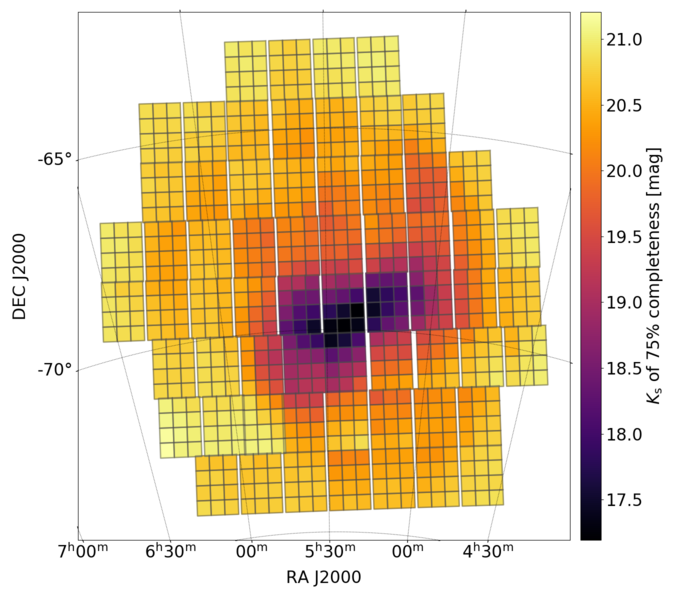}
    \caption{The \textbf{left panel} shows a map of the VMC tiles and subregions considered in this work. Tiles are labelled in cyan, and the inset at the top-right illustrates the numbering of the subregions from G1 to G12, which applies to all tiles. The colour scale indicates the total number of stars detected in both $J$ and \ks\ for each subregion, hence revealing the large scale-structure of the LMC disc and its bar. A few tiles have reduced star counts compared to their neighbours, owing to their particular observing conditions (e.g.\ a worst than usual seeing, the presence of thin cirrus, etc.). The reduced stellar numbers in all subregions G9 are due to the masking we perform to avoid the defective area of VIRCAM detector 16. The \textbf{right panel} shows the same map with a colour scale indicating the \ks\ magnitude at which the completeness falls below 75\%, for stars with $\jks=0$. These maps are obtained from the $J$ and \ks\ photometry, but very similar maps are obtained for the $Y$ and \ks\ photometry. }
    \label{fig:tilesonthesky}
\end{figure*}

\section{The VMC data}
\label{sec:obs}

\subsection{Selected tiles and PSF photometry}

From 2009 to 2018, the LMC was observed in the filters $Y$, $J$ and \ks\ of the VISTA Infra Red CAMera \citep[VIRCAM;][]{vircam}, as part of the VMC survey. For this work we select 63 LMC tiles covering a total area of $\sim\!96$~deg$^2$. Their distribution on the sky is plotted in Fig.~\ref{fig:tilesonthesky}. The larger tile dimension is aligned close to the North--South direction.

For all these tiles, we retrieve the pawprint data from the VISTA Science Archive \citep[][]{vsa}. Image stacking and point spread function (PSF) photometry are performed as described in \citet{rubele15,rubele18}, using the VISTA photometric zeropoints v1.3 \citep[][see also Sect.~\ref{sec:zeropoints} below]{gonzalez18}. 
In the following, we just consider the ``almost-uniformly covered'' section of each tile, i.e. the parts covered at least twice in the tiling of 6 pawprints, with an area of $1.017^\circ\times1.475^\circ$ (out of a total $1.201^\circ\times1.475^\circ$ area observed on the sky). This selection ensures more uniform photometry at a modest cost in terms of total covered area.

The photometric catalogues are then split into 12 subregions per tile for the subsequent SFH analysis. As in \citet{rubele12}, the subregions are numbered from G1 to G12, as illustrated in the left panel of Fig.~\ref{fig:tilesonthesky}. For the sake of brevity, we adopt the following convention in this paper: subregions are referred to as T$t$\_G$g$, where $t$ is an abbreviated number of the LMC tile from VMC, and $g$ is the subregion number from 1 to 12. For instance, according to this scheme the subregion G5 of the tile LMC 3\_2 becomes T32\_G5.

The left panel of Fig.~\ref{fig:tilesonthesky} illustrates the number of stars available per subregion. Some details worth of mention are:
\begin{itemize}
    \item There is a 30\% decrease in the numbers of stars observed in the subregions G9, in the southwestern corner of each tile. This is due to a cut we do in the photometric catalogues, to eliminate the area covered by the top-half of the VIRCAM detector 16, which presents a variable quantum efficiency and hence unreliable photometry \citep[see section 6.1 of][]{vsa}.\footnote{From the $12600\times15500$ pixels of the stacked LMC tiles, the two areas cut from G9 correspond to the intervals with $x=[9700,12600]$, for $y=[0,800]$ and $y=[1870,2740]$.}
    \item Also evident are narrow vertical gaps between adjacent tiles -- regions discarded because observed just once in the tiling of 6 pawprints.
    \item The subregions correspond to projected sizes of about $296~\times322~\mathrm{pc}^{2}$ (in the East-West $\times$ North-South directions) at the LMC distance of $\sim\!50$~kpc.
\end{itemize}
We remark that the total area observed by the VMC survey across the LMC comprises 68 tiles. In this work we use only 63 because for the others, located in the southern part of the LMC, the processing has not yet reached the same level of homogeneity. 

\begin{figure*}
	\includegraphics[width=0.5\columnwidth]{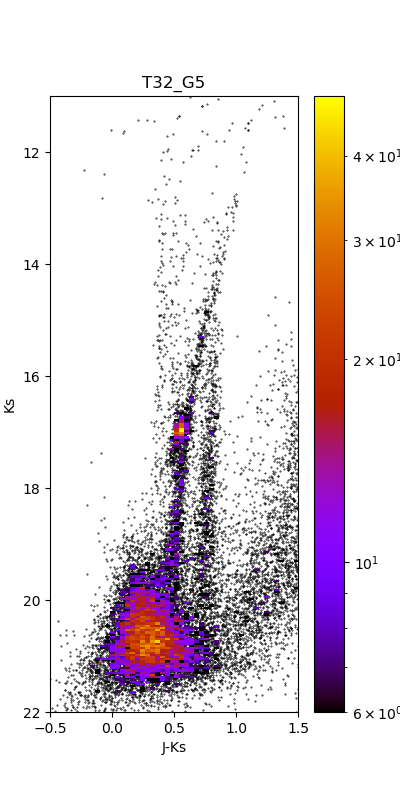}
	\includegraphics[width=0.5\columnwidth]{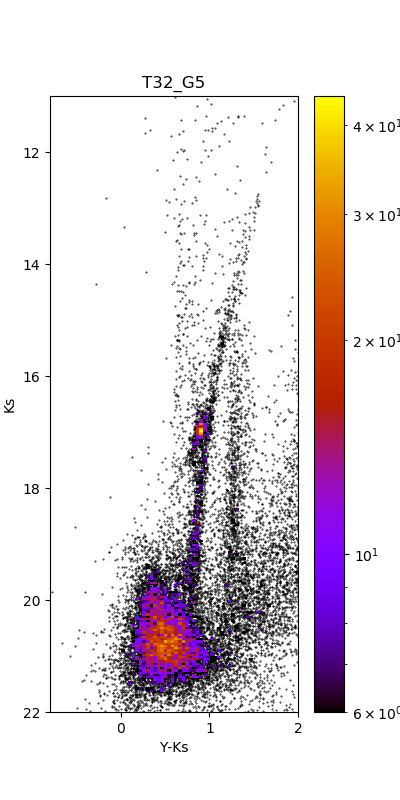}
	\includegraphics[width=0.5\columnwidth]{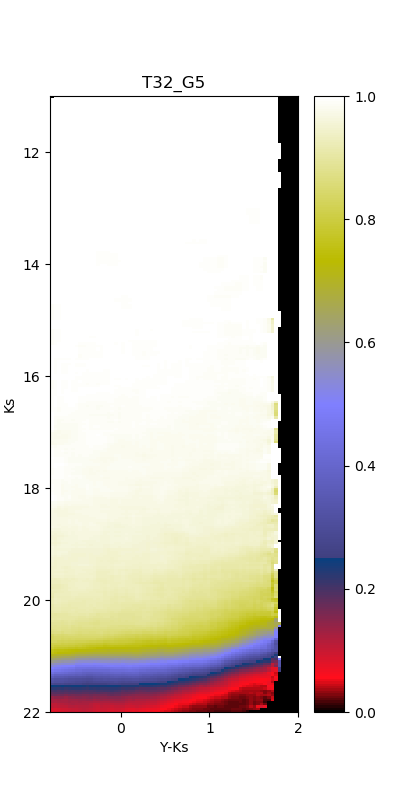}
	\includegraphics[width=0.5\columnwidth]{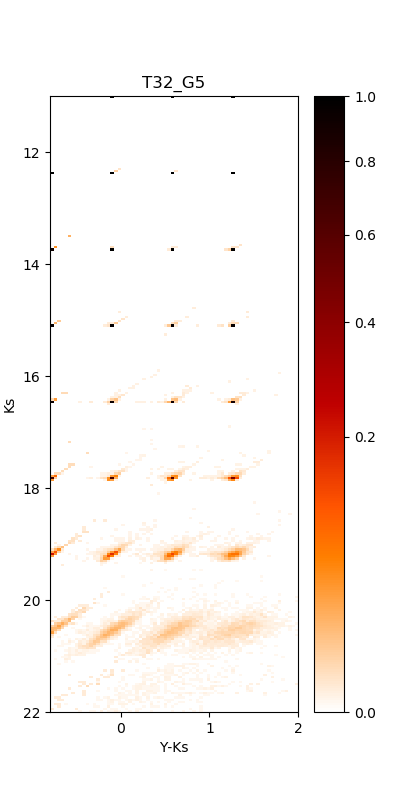}  \\
	\includegraphics[width=0.5\columnwidth]{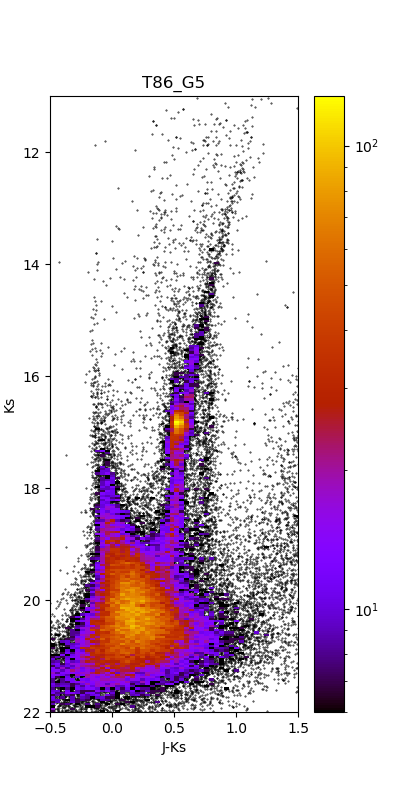}
	\includegraphics[width=0.5\columnwidth]{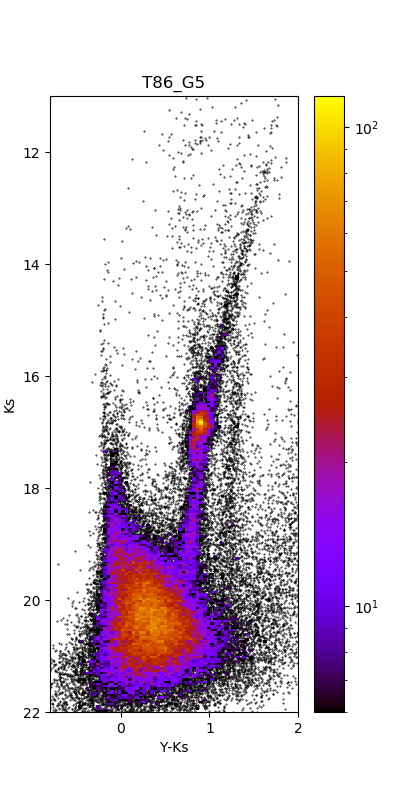}
	\includegraphics[width=0.5\columnwidth]{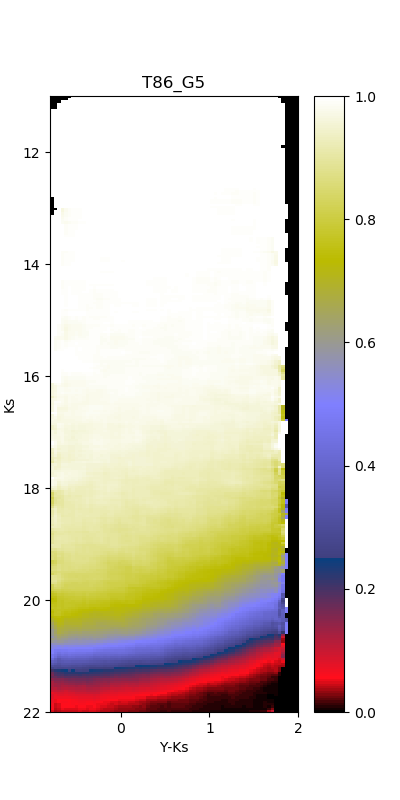}
	\includegraphics[width=0.5\columnwidth]{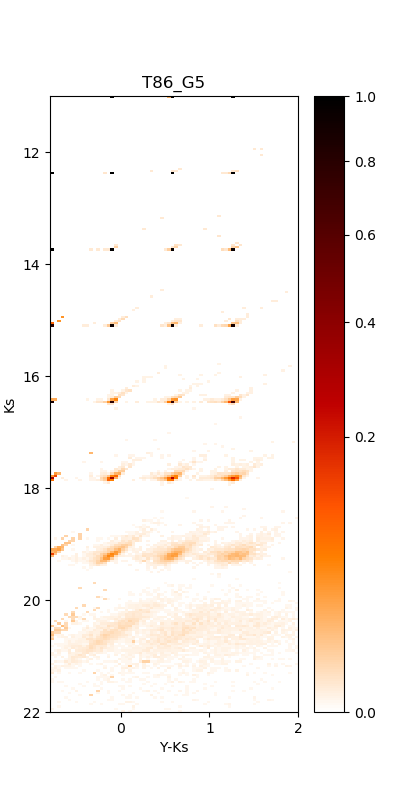}
    \caption{Examples of the observational data used for this work. The \textbf{top row} refers to the subregion T32\_G5, located in the southwestern periphery of the LMC disc. Panels from left to right show the $J\ks$ CMD, the $Y\ks$ CMD, the completeness map for $Y\ks$, and a sample error map in $Y\ks$. The completeness map is filled with zeroes in regions where ASTs were not performed. The sample error map shows how points distributed on a regular grid and with initial density equal to 1, spread in the Hess diagram after convolved with their local error function. Similar diagnostic plots are available for the $J\ks$ dataset; they are not shown just for the sake of brevity -- since there is no dramatic difference between the completeness and error maps in the $Y\ks$ and $J\ks$ datasets (but for the wider colour range covered in the first case). The \textbf{bottom row} presents the same sequence for the subregion T86\_G5, located over the Constellation~III in the northern part of the LMC disc.}
    \label{fig:cmdobs}
\end{figure*}

In Fig.~\ref{fig:cmdobs} we present examples of colour-magnitude diagrams (CMD) derived from these data, for (1) subregion T32\_G5, which is a peripheral, low-density region without signs of star formation younger than $\sim\!1$ Gyr, and (2) T86\_G5, located over the extended Constellation~III in the northern part of the LMC disc, which presents clear signs of recent star formation and a larger stellar density. In the case of T32\_G5, the CMDs present the features expected from intermediate-age and old populations (with ages between 1 and 12.7 Gyr) at the distance of the LMC, in particular the main sequence (MS) at $\ks>19$~mag, the red clump (RC) at $\ks\simeq17$~mag, and the extended red giant branch (RGB) stretching diagonally; moreover, the Milky Way (MW) foreground defines two almost-vertical stripes in the CMDs: a prominent one at $\ks>14$~mag at colours $Y-\ks\simeq1.3$~mag and $J-\ks\simeq0.8$~mag, and a less marked one for all \ks\ and colours $Y-\ks\simeq0.7$~mag and $J-\ks\simeq0.4$~mag. In the case of T86\_G5, additional features coming from young populations become prominent, including a MS extending upwards to $\ks\simeq14$~mag, and the presence of bright core-helium burning stars among the MW features (especially at $\ks\simeq14$~mag, in this case). 

Regions T32\_G5 and T86\_G5 represent most of the range of stellar densities found in the survey: indeed when we rank all subregions by their density of bright ($\ks<18$~mag) and uncrowded stars, T32\_G5 and T86\_G5 are at the 2nd and 71th percentiles of the distribution, respectively. They do not represent the high-density tail of the distribution, which will be discussed later starting from Sect.~\ref{sec:sfhmaps}. Also, we note that even the lowest-dense subregion presents clear LMC features in its CMD, just like in the case of T32\_G5. There are no fields dominated by foreground stars and background galaxies, which could be used as a template to remove the LMC foreground/background.

\subsection{Preparing the initial Hess diagrams}

The SFH analysis essentially consists of reproducing the numbers of stars in many bins across the CMDs. To do that, we start by converting the data into Hess diagrams, which are matrix representations of the stellar density across the CMD, using regularly-spaced colour-magnitude bins. We define Hess diagrams with the following characteristics:
\begin{itemize}
    \item From the $Y\ks$ photometry, we define $275\times70$ bins, for \ks\ between $11$ and $22$~mag with a $0.04$~mag width, and for $\yks$ between $-0.8$ and $2$~mag with $0.04$~mag width;
    \item From the $J\ks$ photometry, we define $275\times50$ bins, with the same limits and spacing for \ks, and for $\jks$ between $-0.5$ and $1.5$~mag with $0.04$~mag width.
\end{itemize}
These limits are wide enough to include the bulk of stars in the observed CMDs (see Fig.~\ref{fig:cmdobs}). The size of the colour-magnitude bins represent a pragmatic choice that will ensure both a good convergence of the SFH-recovery tools (because of the large star counts across the MS, RC and RGB, at least), and a good sensitivity to some astrophysically-interesting quantities such as the distance modulus, mean extinction, and metallicity  (Sect.~\ref{sec:sfh}). Hereafter, we will refer to these two kinds of Hess diagrams as the $J\ks$ and $Y\ks$ cases.

\subsection{Characterizing photometric errors and incompleteness}

As will be discussed below, we also need to assess the error function across the Hess diagrams. It is derived from large sets of artificial star tests (ASTs), namely stars injected into the original VMC images and recovered (or not) with the same PSF photometry pipeline used to derive the VMC catalog. ASTs span the entire sky region being analysed, and are generated across the entire CMD -- although with colours and magnitudes more concentrated around those of the actually-observed stars. We generate a minimum number of $3\times10^5$ ASTs per subregion, increasing this number to $\sim\!2\times10^6$ in some internal LMC areas more affected by crowding. They are injected in multiple runs, at random positions but avoiding self-crowding -- that is, ensuring that the distances between any two artificial stars are larger than the radii involved in the PSF photometry.

Examples of completeness and error maps derived from the ASTs are presented in right panels of Fig.~\ref{fig:cmdobs}, for the $Y\ks$ data of a peripheral (top row) and of a central (bottom row) LMC region. The variation in the completeness between these two cases is evident. In addition, the right panel of Fig.~\ref{fig:tilesonthesky} shows a map of the \ks\ magnitude at which the completeness falls below 75\%, measured at a colour $\jks=0$. This ``incompleteness map'' reveals the dramatic decrease in the photometric depth of the central LMC tiles (6\_6, 6\_5, 6\_4, 5\_5, 5\_6) which largely reflects the increase in the stellar density across the LMC Bar (left panel of Fig.~\ref{fig:tilesonthesky}). Moreover, there are also noticeable tile-to-tile variations in both stellar density and completeness, which reflect the changes in observational conditions during the 9 years of the VMC survey.


\section{The SFH-recovery method}
\label{sec:method}

In previous works deriving the SFH from VMC data \citep{kerber09,rubele12,rubele15,rubele18,pastorelli20} we used a method built around the StarFISH code by \citet{starfish}. This procedure has been completely revised for the present work, with the development of new and more efficient codes to derive both the best-fitting solutions and the confidence intervals of the fitted parameters, and the adoption of a different procedure to build the models. The main changes are described below, while additional details and tests are provided in a recent paper by Dal Tio et al. (submitted).

\subsection{General scheme}

Our final goal is to identify the model $\mathbf{M}$ whose Hess diagram best reproduces the Hess diagram of the observations, $\mathbf{O}$. This model $\mathbf{M}$ is built starting from an ideal model $\mathbf{M}_0$ resulting from a population synthesis code, which contains not only all the information about the LMC populations (distance, extinction and SFH), but also about the MW foreground, that is inevitably present in the data: 
\begin{equation}
\mathbf{M}_0 = \mathbf{M}_0(\mathrm{SFH}, \mu_0, \av, \mathrm{foreground}) 
\label{eq:target_model}
\end{equation}
Many components, denoted by $\mathbf{M}_{0,i}$, are used to compute this ideal model, in the way described in Sect.~\ref{sec:par2tot} below.

$\mathbf{M}$ should also include a simulation of all relevant observational effects, including for instance the photometric errors and incompleteness. This is obtained by convolving all the individual components $\mathbf{M}_{0,i}$ with the error function for every small cell of the Hess diagram $\mathbf{E}(c,m)$:
\begin{equation}
\mathbf{M}_i = \mathbf{M}_{0,i} \ast \mathbf{E}  \,\,\,.
\end{equation}
The error function $\mathbf{E}$ is simply a matrix representation of the completeness and of the colour-magnitude offsets, at every Hess diagram bin, as derived from the ASTs (see Fig.~\ref{fig:cmdobs} above). We typically have a few hundreds of ASTs in a single, well-populated bin of the Hess diagram. Many times more would be advisable for a high-accuracy, noiseless evaluation of the error function, but we cannot achieve that due to the computational cost of the ASTs. We instead take advantage of the slow variation of the error function across the CMDs to average them over boxes $0.24$~mag wide in both magnitude and colour. This is equivalent to multiplying the number of ASTs in every bin by a factor of about 36, hence reducing the shot noise in the derivation of $\mathbf{E}$ by a factor of $\sim\!6$.
 
\subsection{Further culling of the CMDs}
\label{sec:culling}
 
Even though the above-mentioned approach can work well with a lower-than-ideal number of ASTs, on the other hand it brings a limitation to the method: it tells us that the error function cannot be trusted over CMD regions where the completeness (or the photometric errors) varies quickly with either colour or magnitude, within scales comparable to the $0.24~\mathrm{mag}\times0.24~\mathrm{mag}$ boxes within which the error functions are averaged. For this reason, we limit the SFH analyses to regions of the CMD with a completeness higher than 75\% at all colours. Above this limit, we are still in the regime of ``large completeness'' and ``high photon counts'' that ensures very smooth (close to linear) variations in the error function across $\sim0.24$~mag scales in the CMDs. 

In addition, the SFH should be derived from CMD regions that can be reliably simulated with our present-day models -- that is, using stars in the MS, subgiant branch, RGB, RC, and core-helium burning stages of intermediate and large masses. Thermally-pulsing asymptotic giant branch (TP-AGB) stars should better be excluded to avoid a circular argument: even if this evolutionary phase can be well modelled with our codes, the SFH of the Magellanic Clouds was explicitly used in the calibration of the key parameters of present TP-AGB models  \citep[including their lifetimes; see][]{pastorelli19,pastorelli20}. Therefore, there is no sense in using TP-AGB models in the derivation of these same SFHs. Taking these considerations into account, we further limit the SFH analysis to CMD regions with $\ks\!>12$~mag, $-0.5\!<\!\yks\!<\!1.5$~mag, $-0.5\!<\!\jks\!<\!1.0$~mag, so that:
\begin{enumerate}
    \item They include most of the RGB (the RGB tip is located at $\ks\sim\!12.5$~mag; see \citealt{groenewegen19}) and RC stars, even in subregions with a strong reddening. In the latter cases, these features are partially superposed to the vertical feature caused by the MW foreground at $\yks=1.3$~mag and $\jks=0.8$~mag.
    \item At its brightest part, the photometry is not affected by saturation, and is not contaminated by TP-AGB stars in any significant way \citep[see][]{pastorelli20}. 
    \item They exclude most of the faint background galaxies detected at redder colours ($\jks>1$~mag). \label{item:galaxies}
    \item They exclude the bluest part of the CMD, where hardly any star is observed.
\end{enumerate}

We note that the above cuts do not entirely exclude background galaxies from our CMDs. Their complete colour-magnitude distribution, illustrated in figure 6 of \citet{kerber09}, reveals a tail of faint background galaxies extending up to colours as blue as $\jks\sim0.0$~mag. The impact of these galaxies in our method is evaluated in Appendix~\ref{sec:apperrors}.

\begin{table}
    \caption{Parameters used in the TRILEGAL model to compute the MW foreground. The notation is the same as in \citet{pieres20}, while the meaning of each parameter is thoroughly described in \citet{girardi05} and, specifically for the bulge component, in \citet{vanhollebeke09}.
    }
    \label{tab:mwfore}
    \centering
    \begin{tabular}{lcc}
        \toprule
        Component & Parameter identifier & Parameter value \\
        \midrule
        \multirow{2}*{Sun's position}
        & $R_{\odot}$ & 8700~pc \\
        & $z_{\odot}$ & 24.2~pc \\
        \midrule
        \multirow{5}*{Thin disk}
        & $\Sigma_{\odot}^{\mathrm{thin}}$ & 55.41~\Msun\,pc$^{-2}$ \\
        & $h_{R}^{\mathrm{thin}}$ & 2913~pc \\
        & $h_{z,0}^{\mathrm{thin}}$ & 94.7~pc \\
        & $t_{\mathrm{incr}}^{\mathrm{thin}}$ & 5.55~Gyr \\
        & $\alpha$ & 1.67 \\
        \midrule
        \multirow{3}*{Thick disk}
        & $\rho_{\odot}^{\mathrm{thick}}$ & 0.0010~\Msun\,pc$^{-3}$ \\
        & $h_{R}^{\mathrm{thick}}$ & 2394.07~pc \\
        & $h_{z}^{\mathrm{thick}}$ & 800~pc \\
        \midrule
        \multirow{4}*{Halo}
        & $\rho_{\odot}^{\mathrm{halo}}$ & 0.0001~\Msun\,pc$^{-3}$ \\
        & $r_{\mathrm{eff}}$ & 2698.93~pc \\
        & $b_{\mathrm{h}}$ & 0.62 \\
        & $n_{\mathrm{h}}$ & 2.75 \\
        \midrule
        \multirow{4}*{Bulge}
        & $\rho_{\mathrm{GC}}^{\mathrm{bulge}}$ & 406~\Msun\,kpc$^{-3}$ \\
        & $a_{\mathrm{m}}$ & 2500.0~pc \\
        & $a_{0}$ & 95.0~pc \\
        & 1:$\eta$:$\zeta$ & 1:0.68:0.31 \\
        & $\phi_{0}$ & 15\textdegree \\
        \bottomrule
    \end{tabular}
\end{table}

\subsection{The partial and total models}
\label{sec:par2tot}

A total model described by Eq.~\ref{eq:target_model} can be defined as a sum of partial models
\begin{equation}
\mathbf{M} = \mathbf{PM}_0 + \sum_i a_i \mathbf{PM}_i  \,
\end{equation}
where:
\begin{itemize}
    \item[-] $\mathbf{PM}_0$ is the partial model for the Milky Way foreground. It is computed with the calibrated TRILEGAL model \citep{girardi05,girardi16}, whose parameters are listed in Tab.~\ref{tab:mwfore}, for the coordinates and total area under consideration. Foreground extinction is ignored since it is expected to be smaller than 0.2~mag in $A_V$ \citep{subra10}, hence affecting the \ks, $\yks$ and $\jks$ data by less than 0.024, 0.054, and 0.033~mag, respectively. Such changes are much smaller than the MW foreground features observed in VMC data and predicted by TRILEGAL models.
    \item[-] $\mathbf{PM}_i$ are the partial models computed for the LMC, for 16 age bins, and following a given initial age-metallicity relation (AMR) $\feh_0(t)$, and for a reference value of true distance modulus and extinction, namely $\mu_0=18.5$~mag and $A_V=0$~mag\footnote{$\mu_0=18.5$~mag is a classical value for the true distance modulus of the LMC centre \citep{freedman01,pietr09}, differing very little from more recent and accurate determinations \citep[e.g.][]{pietr19}.}. For convenience, all the $\mathbf{PM}_i$ represent stellar populations formed at a constant star formation rate of $\SFRt=1 \Msun\mathrm{yr}^{-1}$, inside the age limits of every bin. Therefore, the coefficients $a_i$ can be directly read as the \SFRt\ in these units.
\end{itemize}

Then, there are small corrections to this model, that allow us to explore small shifts in the global LMC properties, in a fast way:

1) The model can be computed with a given colour-magnitude shift in the Hess diagram, $(\Delta c, \Delta m)$, being applied to all the $\mathbf{PM}_i$ components. This shift is intended to reproduce the shifts caused by changes in the mean reddening and distance modulus of the individual LMC subregions, with respect to the reference values used to compute the PMs. It can also compensate small errors in the photometric zeropoints (although such a compensation is not applied to the MW foreground, as it should be in the case it were really caused by errors in the photometric zeropoints). To improve the efficiency of our algorithms, the $(\Delta c, \Delta m)$ changes are computed only for a limited set of values, which are integer multiples of the resolution in the CMD. For any non-integer multiple, CMDs are computed as a bilinear interpolation of the CMDs for the 4 neighbouring points in $(\Delta c, \Delta m)$ space. This allows us to compute models at runtime, for arbitrary $(\Delta c, \Delta m)$ shifts, with just a factor 4 increase in computing time with respect to the standard  $(\Delta c=0, \Delta m=0)$ case. We typically allow the code to explore intervals of $0.28$~mag in both $\Delta c$ and $\Delta m$.

2) At all ages, partial models are computed for four additional metallicities, differing by $\Delta \feh=(-0.16,-0.08,+0.08,+0.16)$~dex with respect to the reference AMR. This allows us to compute a model for an arbitrary metallicity shift, $\Delta \feh$, using linear interpolation among the two models which bracket its metallicity. Notably, this method allows us to consider small continuous shifts in metallicity by just doubling the computing time, compared to the fixed-AMR case.

These are crude approximations that could be replaced by the actual computation of models at many intermediate values of $(\Delta c, \Delta m, \Delta \feh)$. This, however, would imply a huge increase in computing time, and impose large amounts of computer memory being allocated for increased sets of ``shifted partial models'', effectively prohibiting us the use of Markov chain Monte Carlo (MCMC) methods in our derivation of the SFH solution (cf. Sect.~\ref{sec:mcmc} below).

Moreover, we could decide that the metallicity shifts $\Delta \feh$ assume independent values for every age interval $i$, hence exploring a wide area of the age-metallicity plane. However, changes in the mean AMR of a galaxy field are expected to take place over timescales of Gyr, which excludes large metallicity variations between any two neighbouring age bins in our sequence of young PMs. It should also be noted that the more parameters we use to describe metallicity shifts at different ages, the longer the SFH-recovery process (see Sect.~\ref{sec:mcmc} below) takes to explore the available parameter space. To limit the metallicity variations to a subset of astrophysically-plausible variations, while using few parameters to describe them, we define two $\Delta \feh$ coefficients at the extreme youngest and oldest ages: $\Delta\feh_1$ and $\Delta\feh_2$. $\Delta\feh$ values for all ages are then computed as a simple linear interpolation between these two extremes, with linear age, $t_i$, as the interpolation parameter. In this way, the value of $\Delta\feh_1$ applies to all young populations, mainly affecting the position of the MS stars, while $\Delta\feh_2$ starts affecting the populations older than a few Gyr, hence mainly affecting the properties of the subgiants and giants in the CMD. As before, this simple scheme represents a compromise in which small variations in our input models are explored, but avoiding changes that would imply large increases in the usage of computing time and computer memory.

\subsection{Computing PMs with TRILEGAL}

For the LMC, we adopt the reference AMR taken from the ``closed-box model $y=0.08$'' from \citet{carrera08}, which compares very well with their age-metallicity data for LMC field stars. The main characteristics of this AMR is the presence of two main periods of chemical enrichment, the first one at early epochs exceeding $\sim\!6$~Gyr, the second one at ages younger than $3$~Gyr. They are separated by a period of slower increase in metallicity that corresponds to a period of reduced cluster and field star formation in the LMC disc \citep[see also][]{carrera11}. In addition to this reference AMR, we assume that the metallicities have an intrinsic Gaussian dispersion of $\sigma=0.05$~dex at all ages.

As for the stellar models, we adopt the PARSEC tracks v1.2S \citep{bressan12,chen14,chen15}, in the form of the isochrones provided by default in the CMD web interface version 3.3\footnote{\url{http://stev.oapd.inaf.it/cgi-bin/cmd_3.3}}. These models have been used in a series of previous works on the VISTA and 2MASS \citep[the Two Micron All-Sky Survey;][]{skrutskie06} near-infrared photometry of Magellanic Cloud populations, generally with good results \citep[e.g.][]{rubele18,Lebzelter_etal18, pastorelli19,pastorelli20,trabucchi19,trabucchi21}. Mass loss between the tip of the RGB and the core-helium burning phase is taken into account with an efficiency of $0.2$ times the value provided by the \citet{reimers} formula \citep[see][]{miglio12}. Since our reference AMR extends to very small metallicities (reaching $\feh=-3.2$~dex at the oldest assumed age of $15$~Gyr), these models also include metal-poor populations rich in horizontal branch stars. The theoretical models are converted into the VISTA photometry, in a Vegamag system, via the transformations described by \citet{YBCpaper}; for the stars we model in this work, they are largely based on grids of model atmospheres and synthetic spectra by \citet{ATLAS9} and \citet{phoenix}.

We adopt the canonical initial mass function (IMF) from \citet[][their eqs. 1 and 2; see also eq. 4.55 of \citealt{kroupa13}]{kroupa01}. Binaries are considered only in the form of detached systems, assuming that 30 per cent of the stars drawn from the IMF have a companion, with a mass ratio between $0.7$ and $1$. This prescription suffices to produce a secondary MS parallel to that caused by single stars, and similar to the one observed in HST photometry of star clusters \cite[see e.g.][]{sollima07}. The IMF is normalized so that its integral from 0.01 to 250~\Msun, taking into account both single and binary systems, produces a total mass of $1$~\Msun.

\begin{table}
	\centering
	\caption{Age bins adopted, with their corresponding metallicity interval.}
	\label{tab:age_bins}
    \begin{tabular}{l|llll}
        \hline
        $i$ & \logtyr\ & $\Delta t$ & ${\feh}_0$ & \\
         &  & (yr) & interval (dex) & \\
        \hline
        1 & 6.6, 6.9 &   $3.96\times10^6$  & $-0.19,-0.19$ &  \\ 
        2 & 6.9, 7.2 &   $7.91\times10^6$  & $-0.19,-0.19$  &  \\ 
        3 & 7.2, 7.5 &   $1.58\times10^7$  & $-0.19,-0.19$  &  \\ 
        4 & 7.5, 7.8 &   $3.15\times10^7$  & $-0.19,-0.19$  &  \\ 
        5 & 7.8, 8.1 &   $6.28\times10^7$  & $-0.19,-0.19$  &  \\ 
        6 & 8.1, 8.4 &   $1.25\times10^8$  & $-0.19,-0.19$  &  \\ 
        7 & 8.4, 8.6 &   $2.50\times10^8$  & $-0.19,-0.19$  &  \\ 
        8 & 8.6, 8.8 &   $2.32\times10^8$  & $-0.19,-0.19$  &  \\ 
        9 & 8.8, 9.0 &   $3.69\times10^8$  & $-0.19,-0.19$  &  \\ 
        10 & 9.0, 9.2 &  $5.85\times10^8$  & $-0.19,-0.25$  &  \\ 
        11 & 9.2, 9.4 &  $9.17\times10^8$  & $-0.25,-0.36$  &  \\ 
        12 & 9.4, 9.6 &  $1.47\times10^9$  & $-0.36,-0.49$  &  \\ 
        13 & 9.6, 9.8 &  $2.33\times10^9$  & $-0.49,-0.60$  &  \\ 
        14 & 9.8, 10.0 &  $3.69\times10^9$  & $-0.60,-0.95$  &  \\ 
        15 & 10.0, 10.1 & $2.59\times10^9$  & $-0.95,-2.07$  &  \\ 
        16 & 10.1, 10.2 & $3.26\times10^9$  & $-2.07,-3.18$  &  \\ 
        \hline
 \end{tabular}
\end{table}

We create PMs for 16 age bins (see Table~\ref{tab:age_bins}), at almost-regular intervals of \logtyr, but with widths that become narrower for older populations: bins are 0.3~dex wide starting at $\logtyr=6.6$, and becoming $0.2$~dex wide after $\logtyr=8.4$. Moreover, after $\logtyr=10.0$ we have two bins $0.1$~dex wide, which have very similar Hess diagrams, but with the oldest age bin having a much more extended horizontal branch. The reason for adopting wider age bins at younger ages is essentially to reduce the errors in the determination of the SFH, as discussed in \citet{kerber09}. 

\subsection{The model likelihood}

Summarising, with these choices, a model is determined by (1) the region coordinates, which define the foreground model $\mathbf{PM}_0$, and (2) a set of variable parameters that define the LMC populations. The latter comprise
\begin{itemize}
\item the coefficients for 16 age bins, $a_i$, representing the mean \SFRt\ in every age interval;
\item a global shift in the CMD in both color and magntiude, $(\Delta c, \Delta m)$, which is intended to reproduce the shifts caused by reddening and changes in distance modulus (with respect to the reference values used to compute PMs), and/or small errors in the photometric zeropoints;
\item metallicity shifts $\Delta\feh_1$ and $\Delta\feh_2$, representing small changes in mean metallicity at the extreme ages, which affect the final AMR at all ages.
\end{itemize}

Given this model, for the model-data comparison we adopt the following definition of likelihood ratio derived from a Poisson distribution:
\begin{equation}
    \lnL = \sum_k \left( O_k-M_k - O_k \ln\frac{O_k}{M_k} \right)
\end{equation}
\citep{vanhollebeke09, dolphin02} where $O_k$ and $M_k$ are the observed and model star counts, respectively, in all CMD bins of index $k$ not masked by our selection criteria of Sect.~\ref{sec:culling}. For all CMD bins in which there is a significant number of observed and model stars, results are similar to half of the classical $\chi^2$ (or Gaussian likelihood ratio), where the standard deviation is given by the square root of the observed star counts \citep[see the discussion in][]{dolphin02}.

\subsection{Finding the best solutions}
\label{sec:mcmc}

The search for the maximum-likelihood model is performed in two steps. The first one is a Nelder-Mead minimization based on the \citet{nr} routine, which quickly adjusts the $a_i$ coefficients while keeping all the other parameters fixed (which is equivalent to assuming null shifts in metallicity, distance and reddening). This approximate solution provides an initial model for a Metropolis-Hastings MCMC \citep{metropolis53}, where all parameters are allowed to vary. As a rule, the MCMC is performed with 500 walkers and in 8000 steps, using a new C code (namely \verb$trifit$; see Dal Tio et al.\ submitted) built according to the guidelines from \citet{hogg18}.

\begin{figure*}
	\includegraphics[trim=0 0.8cm 0 0,clip,width=\textwidth]{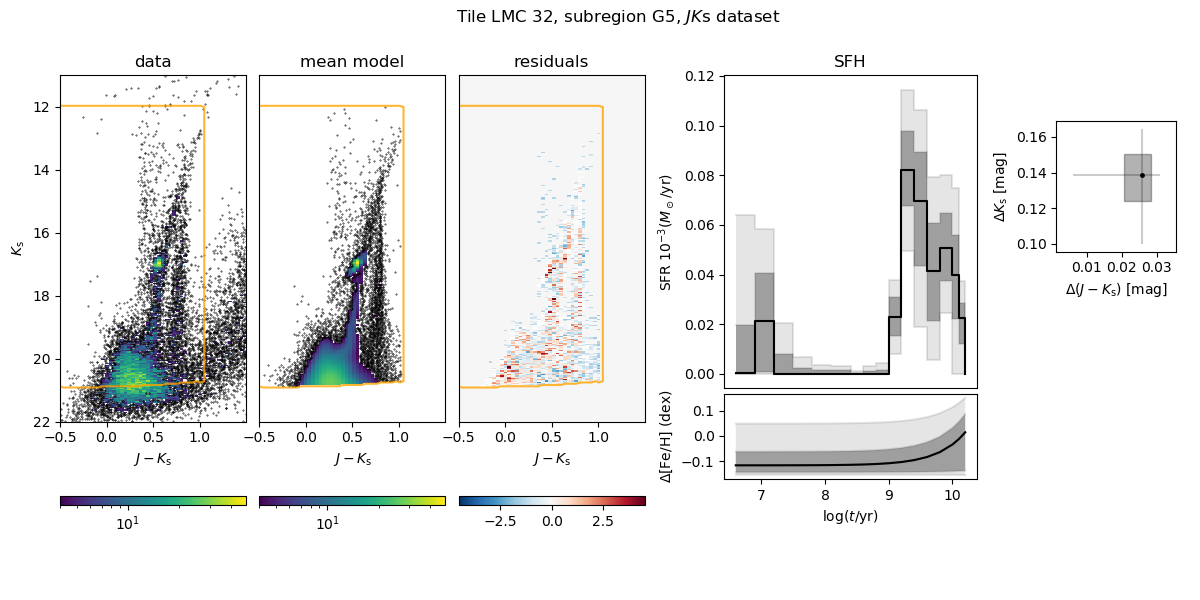}
	\includegraphics[trim=0 0.8cm 0 0,clip,width=\textwidth]{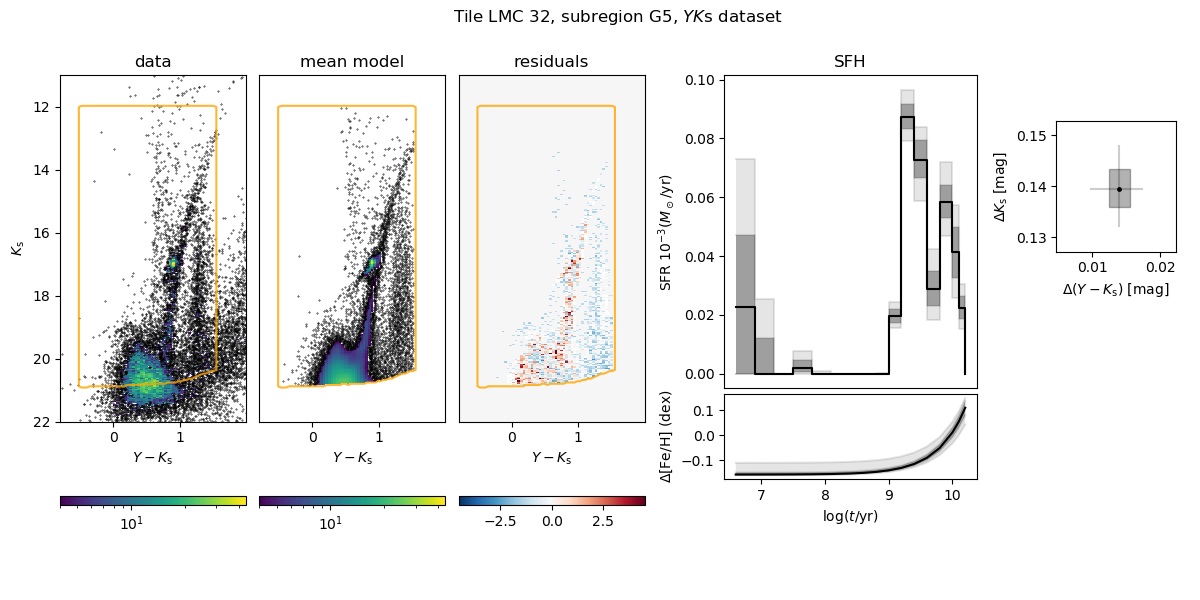}
    \caption{Summary plots presenting the results of fitting the $J\ks$ and $Y\ks$ data (top and bottom rows, respectively) of subregion T32\_G5. The three panels from left to centre show the Hess diagrams of the data, of the mean MCMC solution, and finally the residual between the first two panels, in units of $\sigma$. Note that the models are computed only for the CMD area inside the orange line, which takes into consideration our colour-magnitude cuts and the adopted threshold in completeness. Moreover, the ``mean model'' is represented in an idealised way, with individual stars being randomly simulated in regions of low density, so that it can be plotted in the same way as the actual observations; the actual mean model instead is a Hess diagram containing ample areas with lower-than unity values. The rightmost panels show the results in terms of the fitted parameters, and the astrophysical quantities they represent. These results include the SFH, which is made of (1) the SFR$(t)$ in units of $\Msun\mathrm{yr}^{-1}$ plus (2) the \feh\ differences with respect to our reference values, for which we plot the median values (dark lines), and their 68\% and 95\% confidence intervals (grey and light-grey areas, respectively). Finally, the results include (3) the colour-magnitude shifts with respect to the reference initial value, which are shown in the rightmost panel, with median value (dot) plus 68\% and 95\% confidence intervals (grey box and light-grey error bars, respectively).}
    \label{fig:fitexample-YKs}
\end{figure*}

\begin{figure*}
	\includegraphics[trim=0 0.8cm 0 0,clip,width=\textwidth]{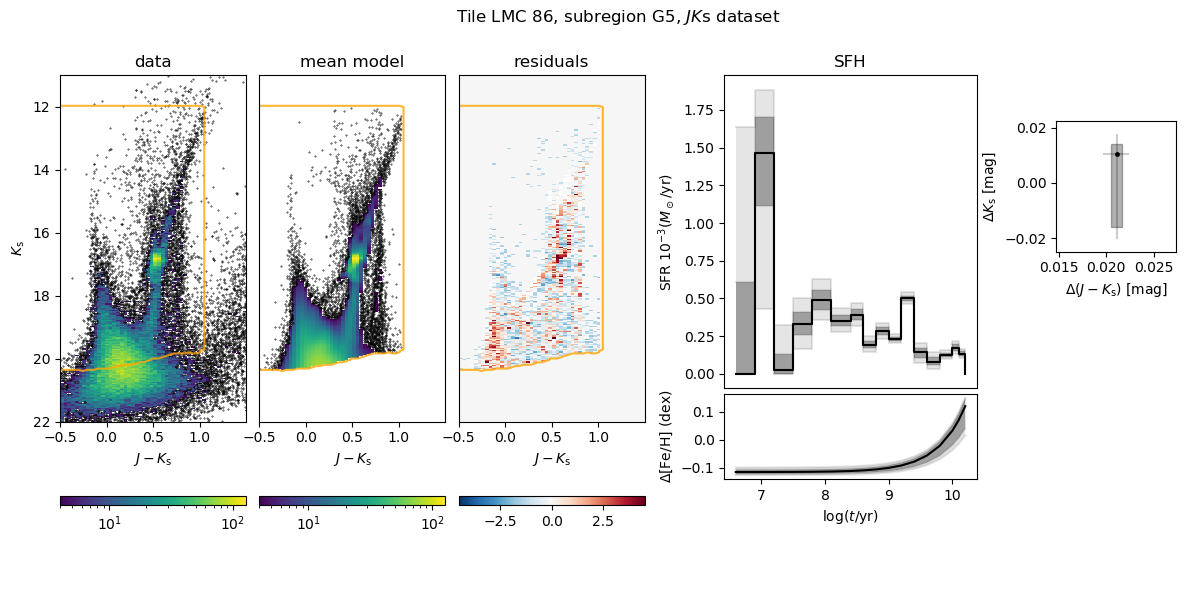}
	\includegraphics[trim=0 0.8cm 0 0,clip,width=\textwidth]{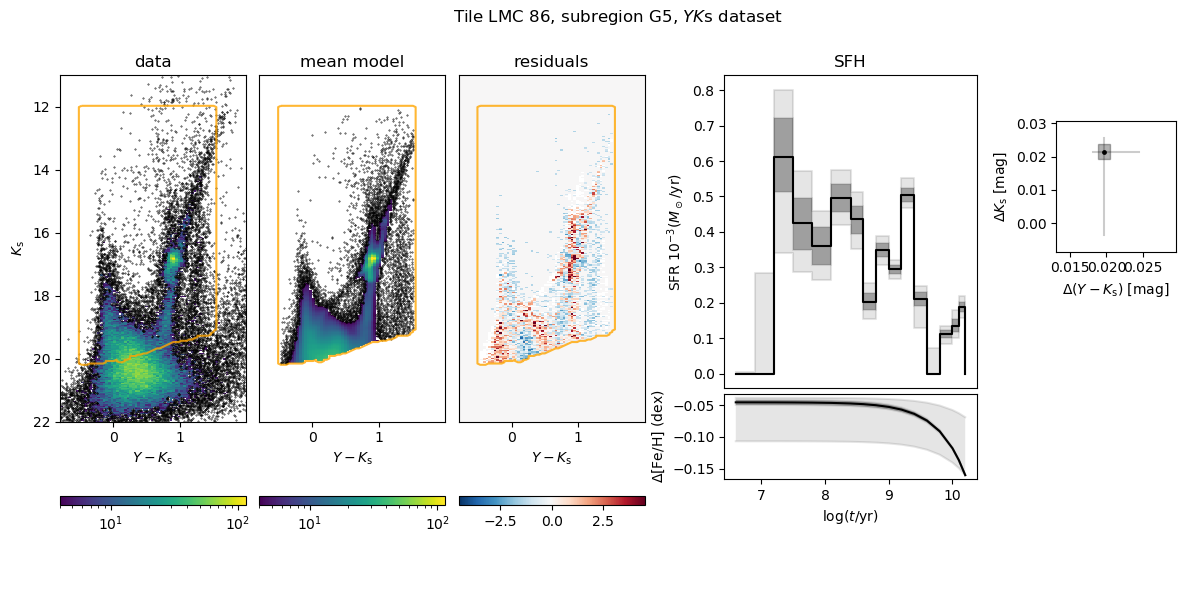}
    \caption{The same as Fig.~\ref{fig:fitexample-YKs}, but for the T86\_G5 subregion.}
    \label{fig:fitexample-JKs}
\end{figure*}

Our best-fit solution is represented by the complete set of final walkers positions. From these, we determine the median, and 68\% and 95\% confidence intervals of all parameters. Appendix~\ref{sec:catalog} describes the machine-readable tables that are available. We also produce the mean Hess diagram from the final walkers, which is then compared to the observed one. Examples of solutions are provided in the summary plots of Figs.~\ref{fig:fitexample-YKs} and \ref{fig:fitexample-JKs}.  

For a few subregions, we ran much longer MCMCs (up to 100\,000 steps), without finding any significant reduction in the final values of $-\lnL$ or changes in the derived parameters. We also tried setting our initial guesses for $a_i$ at different threshold values, in age bins for which the Nelder-Mead step was indicating null values. In all cases the MCMCs converged to the same final set of parameters values, within their 95\% confidence intervals. We therefore consider the minimisation process to be robust.

One caveat is worth of mention at this point: looking at the final distribution of median parameters for all subregions, we notice a concentration of the $(\Delta c, \Delta m)$ values on a grid with an approximate spacing of $0.04~\mathrm{mag}\times0.04$~mag, for both the $J\ks$ and $Y\ks$ datasets. Also the final walkers for every subregion concentrate on a similar grid, but sometimes with sub-concentrations appearing with a different spacing (as small as $\sim\!0.02$~mag). This ``grid effect'' probably results from the adoption of bilinear interpolations to estimate models at intermediate values of $(\Delta c,\Delta m)$. These interpolations are extremely quick to compute and produce the continuous variation of the $\lnL$ values that is required by the MCMC code, but introduce discontinuities in the derivatives of $\lnL$ at the borders of the grid cells. They also produce ``saddles'' in the $\lnL$ values near the central spots of the grid, hence creating regions in $(\Delta c,\Delta m)$ space where the solutions can accumulate.
Eliminating this effect implies either adopting a finer grid of models, or alternative interpolation algorithms, in both cases with a significant cost in terms of computing time. For the moment, we simply prefer to accept these errors, considering them as systematic errors in the derivation of our extinction and distance values (see Sects.~\ref{sec:extinction} and \ref{sec:distance} below).

\section{Results for all subregions}
\label{sec:sfh}

\subsection{The SFH maps}
\label{sec:sfhmaps}

Examples of our solutions, derived independently from the $J\ks$ and $Y\ks$ datasets, are provided in Figs.~\ref{fig:fitexample-YKs} and \ref{fig:fitexample-JKs}. Comparing the different cases one immediately notices that the SFHs derived from $J\ks$ and $Y\ks$ are qualitatively (if not quantitatively) very similar, except for very young ages ($\logtyr<7.2$).

For most of the subregions, the mean model solutions are remarkably similar to the real data in the Hess diagrams. But looking at the map of residuals, some discrepancies appear concentrated in small CMD regions, for instance around the RC. Moreover, the models tend to present the LMC features slightly sharper than in the data, whereas the model MW features sometimes are more spread in colour than the data (especially in \yks). Actually, we find that many of these discrepancies could disappear from our sight if we were to plot the data with colour-magnitude bins twice as large as their present size, or if we were to plot the single model with the maximum likelihood instead of the mean model derived from all final walkers.
These discrepancies could be attributed to a series of problems, going from deficiencies in stellar models to the imperfect simulation of astrophysical effects (e.g. a possible intrinsic dispersion in extinction and distance) and observational effects (e.g. possible inaccuracies in the zero-points, insufficient ASTs, etc.). Investigating these possibilities would be very demanding both computationally and in terms of the effort required to properly assess the effects of each one.

\begin{figure*}
	\includegraphics[width=\textwidth]{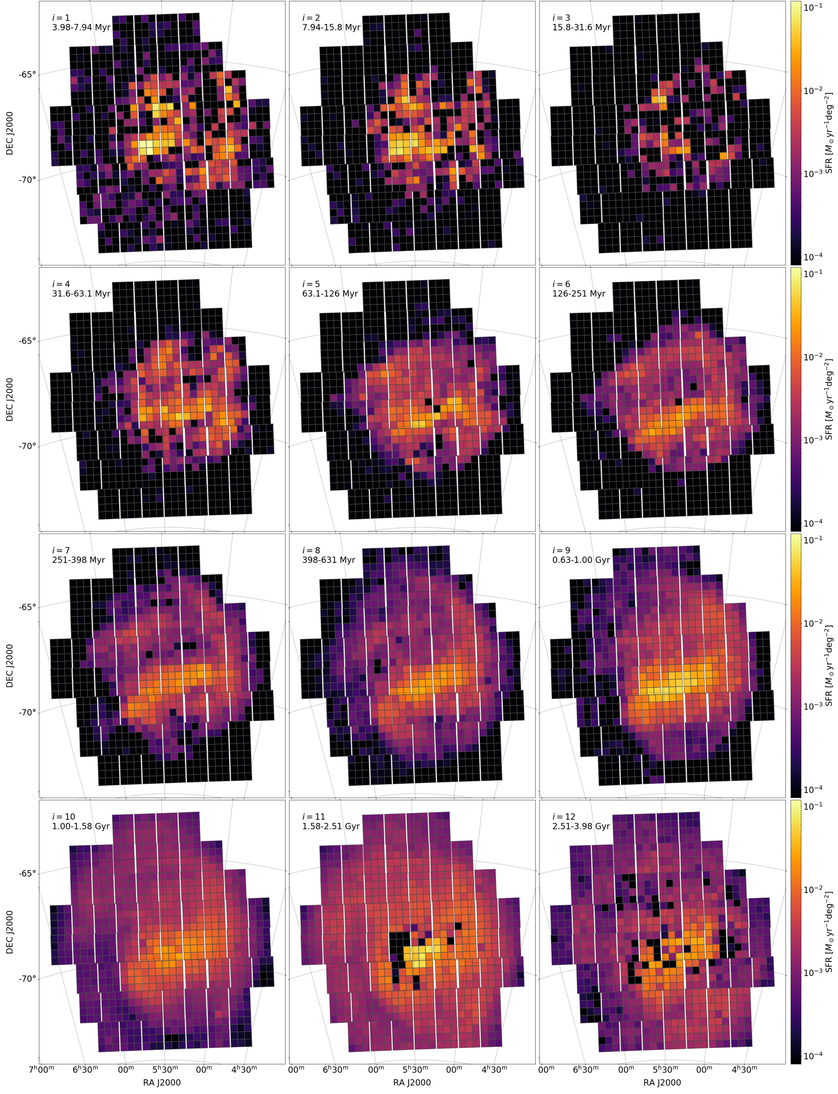}
    \caption{\SFRt\ maps from derived the $J\ks$ data. The 16 panels present the median value of the \SFRt\ in units of $\Msun\mathrm{yr}^{-1}$ divided by the effective area of every subregion, for all our age bins. The colour scale is the same in all panels.}
    \label{fig:SFH_JKs_maps}
\end{figure*}

\begin{figure*}
	\includegraphics[width=\textwidth]{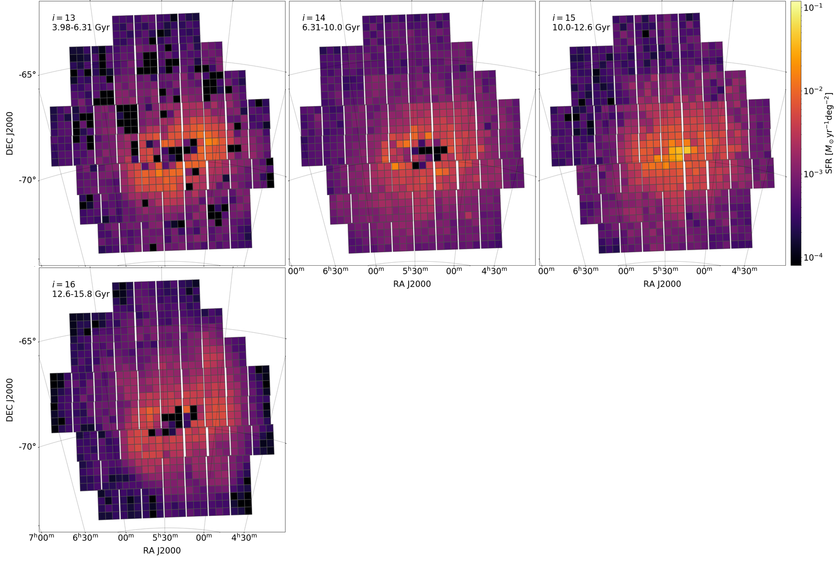}
    \contcaption{.}
\end{figure*}

Let us now take a look at the whole picture obtained combining all the solutions, that is, at the SFH maps of Figs.~\ref{fig:SFH_JKs_maps} and \ref{fig:SFH_YKs_maps}, which present the spatial distribution of the inferred \SFRt\ at all age intervals. A few features are immediately evident in these plots:
\begin{enumerate}
    \item The patchy distribution of \SFRt\ for all ages younger than about $63$~Myr (first row in the figures). At these ages, the \SFRt\ strongly varies from subregion to subregion, and the only large-scale feature is the concentration in a central disc of diameter $\sim\!7$~deg.
    \item At older ages, the \SFRt\ is more smoothly distributed, and becomes more extended as age increases. A notable aspect of these maps is that the \SFRt\ is concentrated on the LMC bar, and in (at least) 3 well-defined spiral arms clearly visible, for example, in the age bin 8, two to the north of the bar and one to the south of it. These configurations persist until ages of at least $1.6$~Gyr -- but with the spiral arms becoming progressively less prominent with increasing age.
    \item For even older ages, the central bar becomes less and less defined, until the \SFRt\ maps eventually reveals a wide, nearly-spheroidal structure distributed over a diameter exceeding $10$~deg, for ages older than $\sim\!1.6$~Gyr.  A clear problem then appears, starting at about the same ages: the maps present a number of holes, corresponding to either isolated subregions, or to groups of neighbouring subregions, where the median \SFRt\ falls to values close to null. These are either subregions where the old \SFRt\ presents very large errors (in practice, we are deriving just an upper limit to the \SFRt), or where the SFH-determination method fails to find astrophysically reasonable solutions.
    The subregions affected are often associated with large amounts of differential 
    extinction (see Sect.~\ref{sec:extinction}) and/or extreme crowding. 
\end{enumerate}
It is probably no coincidence that the latter problem starts at the $1.6$~Gyr ages in which the RGB and RC develop in stellar populations. Starting at that age, the information about the \SFRt\ comes less and less from the dimming turn-off and subgiant branch, and more and more from the RGB and RC regions of the CMD. The RGB and RC are notorious for being very concentrated in the CMD and even more so in NIR CMDs. It is enough to have these two features blurred by differential extinction to lose much of the age information they contain. Interestingly, the \SFRt\ for age bin number 15 appears remarkably smooth in these maps; it is associated with old populations with a short horizontal branch, which do not concentrate in the CMD as much as the RC.

\begin{figure*}
	\includegraphics[width=0.48\textwidth]{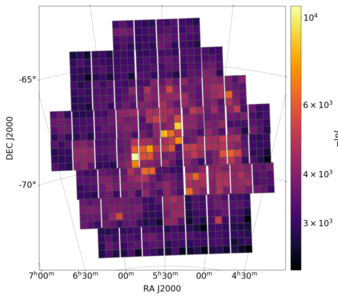}
	\includegraphics[width=0.48\textwidth]{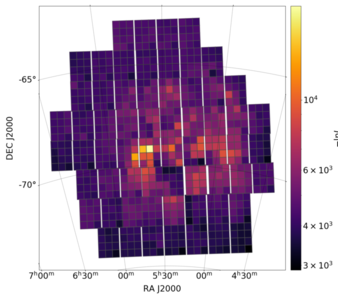}
    \caption{Maps of likelihood, $-\lnL$, for the $J\ks$ (left) and $Y\ks$ (right) solutions.}
    \label{fig:mlnL_maps}
\end{figure*}

Can we tell in advance, just looking at the final likelihood of the fit, what are the subregions for which the old \SFRt\ is unreliable? This is not so easy. However, the maps of likelihood, presented in Fig.~\ref{fig:mlnL_maps}, do indicate several regions of higher-than-average $-\lnL$, and many of them (but not all) coincide with holes in the old-\SFRt\ maps.

Another aspect to consider is that the holes in the \SFRt\ maps at certain ages are, at least partially, compensated by a higher \SFRt\ at neighbouring age intervals. This is because the fitting algorithm will always try to produce RGB and RC stars in appropriate numbers to fit the CMD. Whenever the extinction is moderate, the stars more suitable to produce these appropriate numbers will likely be found in nearby age intervals -- but not at age intervals with $<\!1.5$ Gyr, which have dramatically different CMD features.

These problems could probably be solved by using additional data -- for instance detailed maps of internal reddening, or CMDs from optical data being analysed simultaneously with the VMC data -- in the CMD-fitting method. Such options will be explored in subsequent papers.

\subsection{Consistency checks on the SFR$(t)$}

\begin{figure}
	\includegraphics[width=\columnwidth]{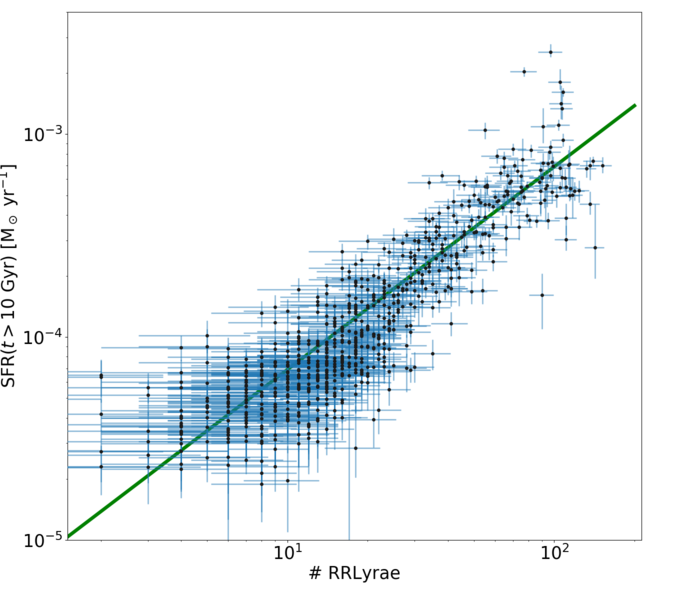}
    \caption{Comparison between the \SFRt\ at very old ages, derived from $J\ks$ data, with the number of RR Lyrae in each subregion. The error bars correspond to the 68 per cent confidence interval of the SFR, and to the square root of star counts. The green line presents the mean relation derived assuming a perfect proportionality between these quantities: its slope is of $1.45\times10^{5}$ RR Lyrae per unit $\Msun\mathrm{yr}^{-1}$.}
    \label{fig:rrl_maps}
\end{figure}

As a consistency check, we compare our maps with the distribution of stars expected to be contained into small age intervals. The first case is offered by RR Lyrae, which were formed $\gtrsim\!10$~Gyr ago, and preferably at low metallicities. Fig.~\ref{fig:rrl_maps} compares the mean \SFRt\ for ages $>\!10$~Gyr as obtained from the $J\ks$ data (see the two last panels of Fig.~\ref{fig:SFH_JKs_maps}), with the absolute number of RR Lyrae in each subregion from the catalog by \citet{cusano21}. Additional plots regarding the spatial distribution of the RR Lyrae, and the \SFRt\ derived from $Y\ks$, are presented in Fig.~\ref{fig:rrl_addmaps}. It is evident that the spatial scale of the RR Lyrae and old \SFRt\ distributions are very similar.
The mean proportionality constant between these quantities is of $1.58\times10^{5}$ and $1.45\times10^{5}$ RR Lyrae ${\Msun}^{-1}\mathrm{yr}$ for the $Y\ks$ and the $J\ks$ maps, respectively. Of course, this relation is influenced by a few of the central subregions, which have high but potentially incomplete numbers of RR Lyrae, and whose \SFRt\ estimates are more affected by crowding and differential reddening. It is also evident that the points in the diagram are more spread than expected from the formal error bars. Overall, this plot is reassuring (because there is a clear correlation), but also indicates that there is space for improvement in the determination of the old \SFRt.

\begin{figure}
	\includegraphics[width=\columnwidth]{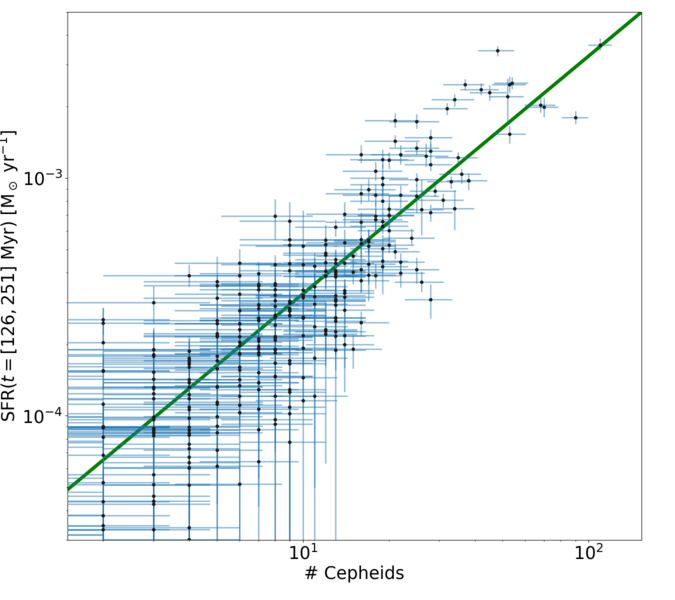}
    \caption{Comparison between the \SFRt\ at ages between $126$ and $251$~Myr, derived from the $J\ks$ data, with the number of Cepheids in each subregion. The error bars correspond to the 68 per cent confidence interval of the SFR, and to the square root of star counts. The green line presents the mean relation derived assuming a perfect proportionality between these quantities: its slope is of $3.07\times10^{4}$ Cepheids per unit $\Msun\mathrm{yr}^{-1}$. }
    \label{fig:cepheids_maps}
\end{figure}

Figure \ref{fig:cepheids_maps} presents the same kind of comparison for the classical Cepheids from the catalogue by Ripepi et al. (in preparation). Their distribution is compared to the \SFRt\ at the 6th age bin, corresponding to ages between $126$ and $251$~Myr. This is the age interval at which the blue extremity of the ``blue loop'' of core-helium burning stars transits from the red to the blue of the instability strip, producing the maximum numbers of Cepheids for a given SFR. Additional plots regarding the spatial distribution of the Cepheids, and the \SFRt\ derived from $Y\ks$, are presented in Fig.~\ref{fig:cepheids_addmaps}. Also in this case, there is an evident correlation between the number of Cepheids and our derived \SFRt. The correlation is somewhat noisy and may be affected by the strong differential extinction of some central LMC subregions. The mean proportionality constant is $3.03\times10^{4}$ and $3.07\times10^{4}$ Cepheids ${\Msun}^{-1}\mathrm{yr}$, for the $Y\ks$ and the $J\ks$ maps, respectively.

\subsection{Correction to photometric zeropoints}
\label{sec:zeropoints}

\begin{figure*}
	\includegraphics[width=0.33\textwidth]{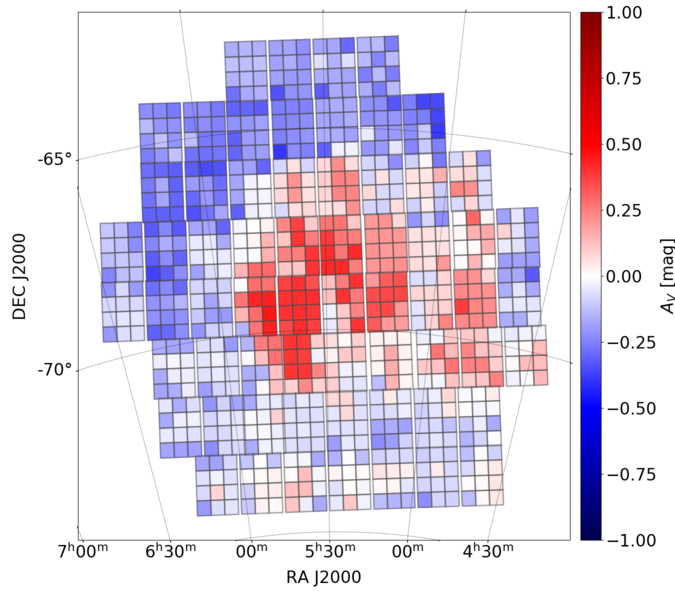}
	\includegraphics[width=0.33\textwidth]{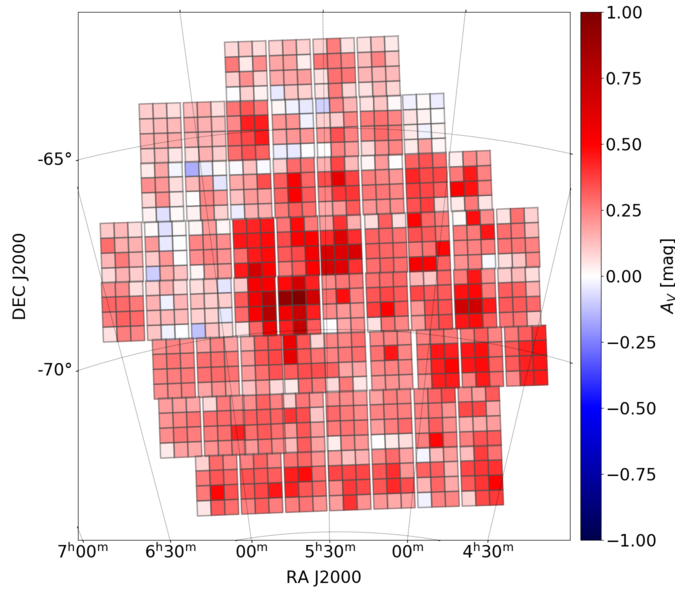}
	\includegraphics[width=0.33\textwidth]{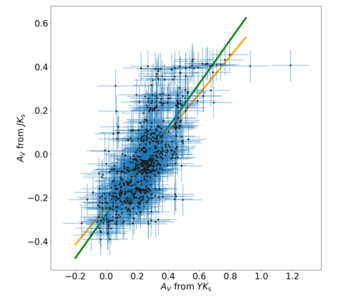}
    \caption{$V$-band extinction maps derived from $Y\ks$ (left panel) and $J\ks$ (central panel) data. The right panel shows the relationship between them, with two fitted relations: a systematic shift of $A_V^{YK_\mathrm{s}}=A_V^{JK_\mathrm{s}}-0.28$~mag (green), or a best-fit line of $A_V^{YK_\mathrm{s}}=0.86\,A_V^{JK_\mathrm{s}}-0.24$ (orange).
    } 
    \label{fig:extinction_maps}
\end{figure*}

Our PSF photometry is based on the v1.3 calibration of VISTA data. When our analysis of the SFH was well underway, we measured the offsets between this photometry and the more accurate v1.5 calibration from \citet{gonzalez18}, on a tile-to-tile basis. They imply corrections to the PSF photometry whose mean and r.m.s. values, for our 63 tiles, are
\begin{eqnarray}
\Delta \mathrm{zp}(\ks) &=& 0.000\pm0.017~\mathrm{mag} \nonumber \\ 
\Delta \mathrm{zp}(\yks) &=& -0.026\pm0.026~\mathrm{mag}  \\
\Delta \mathrm{zp}(\jks) &=& 0.026\pm0.022~\mathrm{mag}  \nonumber 
\end{eqnarray}
These corrections are typically smaller than our adopted CMD resolution, and much smaller than the total shifts in colour and magnitude that we explore during the CMD fitting. Therefore, instead of using them to correct the input photometry, these zeropoint shifts are added, a posteriori, as small corrections to the $(\Delta c, \Delta m)$ derived in our CMD fitting. Then we can define the following ``magnitude and colour excesses'':
\begin{eqnarray}
E(\ks) &=& \Delta m + \Delta \mathrm{zp}(K_\mathrm{s}) \nonumber \\
E(\yks) &=& \Delta c^{YK_\mathrm{s}} + \Delta \mathrm{zp}(Y)-\Delta \mathrm{zp}(K_\mathrm{s}) \label{eq:excesses} \\
E(\jks) &=& \Delta c^{JK_\mathrm{s}} + \Delta \mathrm{zp}(J)-\Delta \mathrm{zp}(K_\mathrm{s}) \nonumber 
\end{eqnarray}
which are necessary to explore the extinctions and distances across the LMC. 

\subsection{Extinction maps}
\label{sec:extinction}

Assuming a \citet{cardelli} extinction law, extinction coefficients of an average dwarf star (the Sun) at the limit of low extinction, are
\begin{equation}
(A_Y,A_J,A_{K_\mathrm{s}}) = (0.391, 0.288, 0.120) \,A_V   \,.
\label{eq:extcoef}
\end{equation}
\citep[see][]{girardi08,rubele12}. These coefficients change little for stars of different effective temperature and surface gravity; for instance they decrease by less than $2$ per cent for cool RGB stars of $\Teff=3500$~K, $\logg[\mathrm{cm\,s}^{-2}]=0.5$. Slightly larger are the variations between coefficients published by different authors, due to the way filter transmission and extinction curves are interpolated and convolved, and to changes in the exact definition of ``low extinction''\footnote{For instance, \citet{YBCpaper} recently derives extinction coefficients smaller by a few percent, i.e.\ $(0.369,0.272,0.118)$, using the same \citet{cardelli} extinction curve.}. Such variations are inconsequential for this work, and they are similar to the differences among coefficients derived from different extinction curves in the literature.

Our CMD-fitting solutions provide the colour excesses (eq.~\ref{eq:excesses}) which we equal to the colour excess caused by interstellar dust. They can be converted into the total extinction with:
\begin{eqnarray}
A_V &=& 3.69\times E(Y-\ks) \label{eq:extinction}\\
A_V &=& 5.95\times E(J-\ks) \nonumber 
\end{eqnarray}
providing two independent values of \av.
They are plotted in the form of extinction maps in Fig.~\ref{fig:extinction_maps}. We choose a representation using $A_V$ because it can be easily compared with the many extinction maps already present in the literature. Some readers could prefer a representation using $A_{K_\mathrm{s}}$, which is easily obtainable by multiplying the colour scale of Fig.~\ref{fig:extinction_maps} by $0.12$.

As for the errors in \av, they derive directly from the errors in the colour excess in eq.~\ref{eq:extinction}. They have two components, which we add in quadrature: the first is the 68\% confidence intervals of $\Delta c$ provided by the MCMC method. The second is half the stepsize in colour of our Hess diagrams, that is, $0.02$~mag. This latter is assumed as a minimum error, necessary to take into account the grid effect we find in our final $\Delta c$ values (see Sect.~\ref{sec:mcmc}).

Negative extinction values do occur in Fig.~\ref{fig:extinction_maps}, and by itself they are indicating that factors other than extinction are contributing to the colour shifts. Likely candidates are offsets in the evolutionary models, which would affect all tiles, and additional (unidentified) offsets in the photometric zeropoints, whose effect could also appear on a tile-to-tile basis. Indeed, we find that:
\begin{itemize}
    \item The $A_V$ map derived from $J\ks$ has just a few peripheral regions with slightly negative values, down to $A_V=-0.2$~mag. Positive values extend up to $A_V=+1.2$~mag.
    \item In the $A_V$ map derived from $Y\ks$, instead, nearly half of the subregions have negative values. The $A_V$ values extend all the way from $-0.4$ to $+0.45$~mag.
    \item There is a correlation between the two $A_V$ values, with a weighted least-squares fit producing either a systematic shift of $A_V^{YK_\mathrm{s}}=A_V^{JK_\mathrm{s}}-0.28$~mag, or a best-fit line of $A_V^{YK_\mathrm{s}}=0.87\,A_V^{JK_\mathrm{s}}-0.24$~mag.
\end{itemize}

The most natural interpretation of these results is that either there is an error in the extinction coefficients applied to the different filters, or there is a systematic error affecting the photometry, \textit{either in the data or in the theoretical models}. In this regard, we note that (1) our adopted extinction coefficients (eq.~\ref{eq:extcoef}) in the VISTA $J\ks$ passbands agree with those determined for 2MASS $J\ks$ passbands in a series of empirical methods \citep{indebetouw05,zasowski09,wang14,schlafly16}, (2) the VMC data is strictly calibrated using 2MASS data, which does not have a $Y$ filter, and finally (3) the model fitting produces reasonable $A_V$ values when only $J\ks$ data are used. All these factors lead us to conclude that the problem is more likely associated with the $Y$ filter, than with the $J\ks$ ones. Indeed, all results would become more reasonable if the $Y$-band photometry was shifted by $+0.07$~mag, or alternatively, if all model $Y$-band magnitudes were shifted by $-0.07$~mag.

\begin{figure*}
	\includegraphics[trim=0 0.8cm 0 0,clip,width=\textwidth]{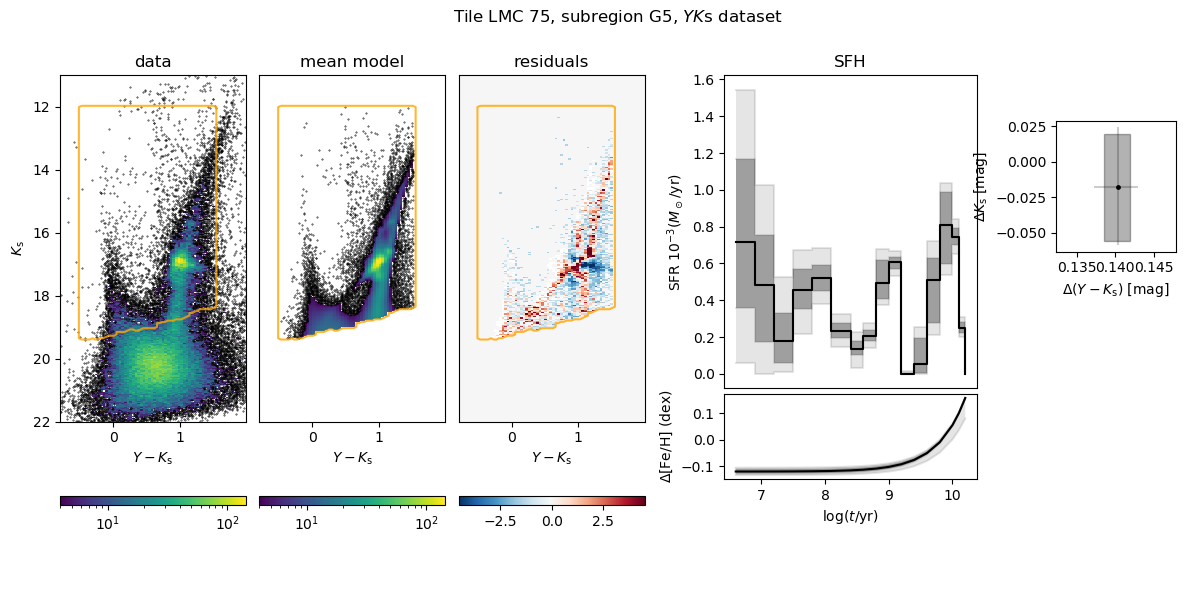}
    \caption{The same as Fig.~\ref{fig:fitexample-JKs}, but for the $Y\ks$ data of the highly-extincted subregion T75\_G5.}
    \label{fig:fitexample-highextinction}
\end{figure*}

The range of parameters adopted (Sect.~\ref{sec:par2tot}) allows us to explore $A_V$ values in the approximate ranges $[-0.45,+0.59]$~mag for $Y\ks$ data, and $[-0.48,+1.20]$~mag for $J\ks$ data. In the case of $Y\ks$ data, about 20 subregions reached values close to the upper \av\ limit, but only in central regions severely affected by crowding, which will be eliminated from most of our discussion further below.

In some subregions, VMC photometry provides significant evidence for a large internal spread in the extinction values. An example is provided in Fig.~\ref{fig:fitexample-highextinction}, for the subregion T75\_G5. Its Hess diagram presents an extended tail of reddened RC stars, rather than a simple shift of the RC to redder colours and fainter magnitudes. They are indicating that extinction is rapidly varying across the subregion, and possibly also varying in depth across the LMC disc. Other examples are given in \citet{tatton13}, who better illustrate the spatial scale at which such extended RC tails appear. Our present models are not built to reproduce this kind of situation. Indeed, the mean models illustrated in Fig.~\ref{fig:fitexample-highextinction} fail to fit the shape of the RC and present high residuals in other parts of the CMD as well.
The unsuitability of this model fit is evidenced by the increased $-\lnL$ in the likelihood maps of Fig.~\ref{fig:mlnL_maps}. Other examples of subregions with highly-non uniform extinction are T85\_G7, T75\_G10 \citep[rich in pre-main sequence stars;][]{zivkov18} and T66\_G7 \citep[containing 30~Dor;][]{tatton13}, all standing out for their higher $-\lnL$ in Fig.~\ref{fig:mlnL_maps}. Substantial (and very time-consuming) changes in the analysis are planned to improve the model fitting in these regions of the LMC disc. 

In addition, visual inspection of Fig.~\ref{fig:extinction_maps} suggests that there are a few \textit{entire tiles} with deviant values of $A_V$. An example is tile 7\_7 in the $J\ks$ extinction map, for which $A_V$ values are about $\sim\!0.2$~mag higher than in neighbouring tiles. These cases likely represent either degraded observing conditions or non-detected errors in the zeropoints of a tile.

Considering the above-mentioned problems, our extinction maps are less detailed than many other extinction maps derived for the LMC.  
However, we note that our high-extinction regions coincide with those derived by many different authors using completely independent data and methods \citep[see for instance][]{zaritsky04, furuta19}.

\subsection{Distance maps}
\label{sec:distance}

\begin{figure*}
	\includegraphics[width=0.33\textwidth]{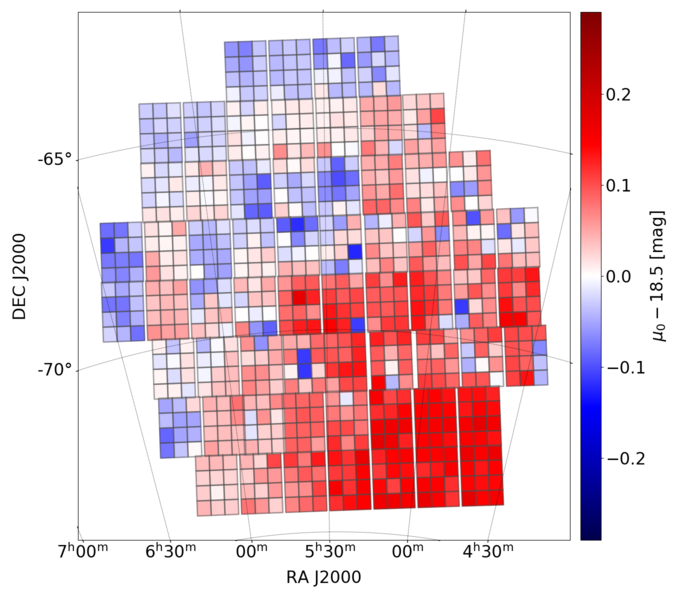}
	\includegraphics[width=0.33\textwidth]{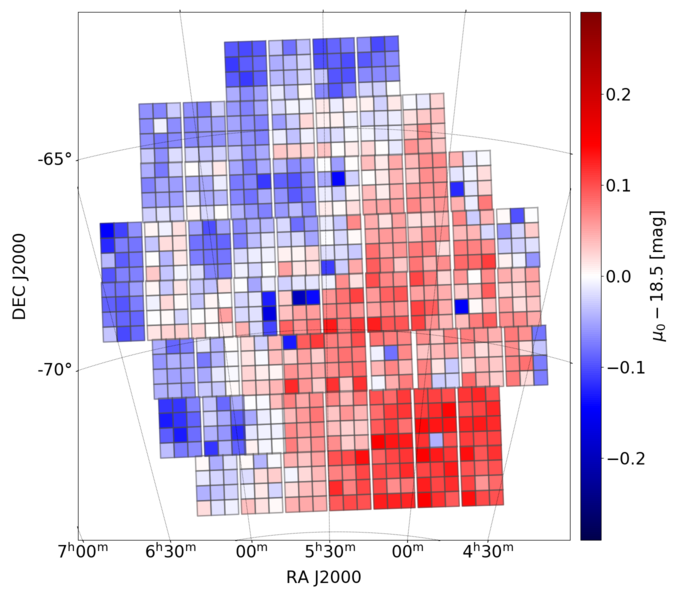}
	\includegraphics[width=0.33\textwidth]{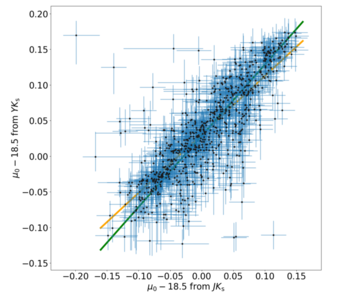}
    \caption{From left to right, we have the true distance moduli maps derived from $Y\ks$ and $J\ks$, and their relationship (as in Fig.~\ref{fig:extinction_maps}). The two fitted relations are: a systematic shift of $\mu_0^{YK_\mathrm{s}}=\mu_0^{JK_\mathrm{s}}+0.03$~mag (green), or a best-fit line of $\mu_0^{YK_\mathrm{s}}-18.5=0.82\,(\mu_0^{JK_\mathrm{s}}-18.5)+0.03$~mag (orange).
    }
    \label{fig:distmod_maps}
\end{figure*}

Similarly to extinction, we explore shifts in magnitude \ks\ that can be interpreted as distance variations across the LMC. True distance moduli can be computed with
\begin{equation}
\mu_0 = 18.50 + E(\ks) - A_{K_\mathrm{s}} 
\label{eq:mu0}
\end{equation}
from both the $J\ks$ and $Y\ks$ cases. Results are plotted in Fig.~\ref{fig:distmod_maps}.

Errors in $\mu_0$ come from the 68\% confidence intervals indicated by the MCMC results, plus a minimum $0.02$~mag error added in quadrature to take into account the grid effect in the $\Delta m$ values (similarly to Sect.~\ref{sec:extinction}). Therefore, these maps incorporate the already-mentioned issues in the extinction maps, but fortunately enough this is a minor problem in the \ks\ band. Despite of the negative extinctions values derived from the $Y\ks$ data, the true distance moduli obtained from the $Y\ks$ and $J\ks$ cases have similar values, with a weighted least-squares fit indicating a systematic shift of just $0.03$~mag between them.
Much more worrying, instead, are the irregularities, on a tile-to-tile basis, present in the distance maps.

Anyway, a clear picture results from these distance maps: southwestern LMC regions are systematically farther away than northeastern regions. This is in agreement with all literature to date, and will be explored further in Sect.~\ref{sec:geometry} below.

\subsection{Metallicity shifts}
\label{sec:metallicity}

Near-infrared photometry is in general less sensitive to metallicity than optical photometry, especially when distance modulus and extinction are considered as free adjustable parameters, as in our case. Anyway there is some dependency on metallicity encoded in VMC data, coming from subtle changes in the mean slope of the RGB, in the position and shape of the RC, and in the mean colour (compared to the RGB) and slope of the MS. This information is partially recovered by our MCMC code, in the form of the metallicity shifts with respect to the initially-assumed AMR from \citet{carrera08}, $(\Delta\feh_1,\Delta\feh_2)$, at both very young and very old ages.

The resulting metallicity shifts present a lot of scatter, and no evidence of large-scale metallicity gradients in the LMC region presently explored. However, they clearly indicate that young LMC populations are better represented by metallicities slightly lower than those in the reference AMR from \citet{carrera08}. Median values for the $(\Delta\feh_1,\Delta\feh_2)$ coefficients are $(-0.084,0.049)$~dex for the $Y\ks$ data, and $(-0.092,0.004)$~dex for the $J\ks$ data. Work is ongoing (Choudhury et al., in preparation) to explore the LMC metallicity trends from VMC data in a more systematic way.

\section{Discussion} 
\label{sec:discu}

Now that we have large-scale maps of \SFRt\ with different degrees of reliability across the LMC disc, and additional information about large-scale changes in properties such as extinction and distance, we discuss ways in which we can better interpret and improve these results, while still adopting the present data, models and algorithms.

\begin{figure}
	\includegraphics[trim=0 0 2.5cm 0,clip,width=\columnwidth]{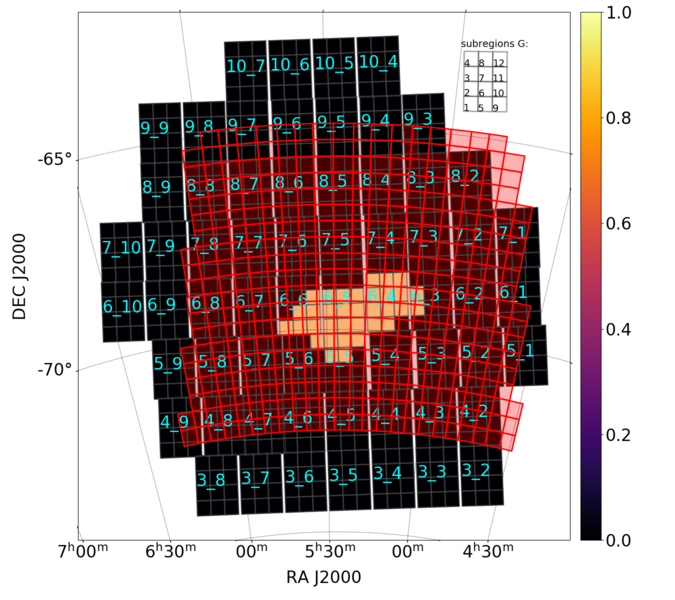}
    \caption{Comparison between the sky areas covered by our SFH maps (dark and yellow areas) and those of \citetalias{HZ09} (red/pink areas). The yellow areas are those more affected by crowding $(\ks(\mathrm{75\%completeness})<18.5$~mag), which are excluded from some of our comparisons.}
    \label{fig:compareas}
\end{figure}
 
\subsection{The LMC geometry}
\label{sec:geometry}

A first-order description of the LMC geometry is that the young and intermediate-age populations are on an inclined thin disc, as suggested by distance indicators such as Cepheids \citep[e.g.][]{nikolaev04} and RC stars \citep[e.g.][]{olsen02}, while the very old populations have a thicker and more spheroidal distribution, as indicated by the RR Lyrae \citep[e.g.][]{subramaniam06}. Assuming that our method is efficiently measuring the mean distance of every subregion, we can fit an inclined disc to our sets of true distance moduli, similarly to what was done by the above-mentioned authors, and updating a similar fit which was done by \citet{rubele12} using just 4 tiles of VMC data. To fit the plane, we proceed as follows:
\begin{itemize}
    \item We fix the coordinates of the LMC centre, $(\alpha_\mathrm{c},\delta_\mathrm{c})$, adopting one of the several possible choices in the literature (see Table~\ref{tab:plane}).
    \item Every subregion is located in Cartesian space by using its central coordinates and the median value of $\mu_0$ determined from our fit (eq.~\ref{eq:mu0}), together with its error. 
    \item The best-fitting plane is determined by looking for the heliocentric distance to the LMC centre, $R_0$, the disc inclination on the plane of the sky, $i$ (where $i=0^{\circ}$ means a face-on disc), and the position angle of the line of nodes, $\theta_0$, that minimizes the distance residuals to the plane. 
    \item The initial guess of $(R_0,i,\theta_0)$ is taken from literature values, and it is let to evolve via the MCMC code \textit{emcee} \citep{emcee}.
\end{itemize}
To limit the discussion to the most reliable subregions, we also apply the following cuts to the set of subregions:
\begin{itemize}
    \item $\ks(\mathrm{75\%~completeness})> 18.5$~mag
    \item $-\lnL<6000$
\end{itemize}
They eliminate the central LMC from the fit, including a large fraction of its Bar (see Fig.~\ref{fig:compareas}).

\begin{table*}
    \centering
    \caption{Results for the fitting of the LMC plane using the distances to many subregions.}
     \label{tab:plane} 
    \begin{tabular}{lllllll}
    \hline
    case & $\alpha_\mathrm{c, J2000}$ & $\delta_\mathrm{c, J2000}$ & $i$ & $\theta_0$ & $R_0$ & comments\\
         & (deg) & (deg) & (deg) & (deg) & (kpc)$^1$ & \\
    \hline
    \citet{vandermarel01}$^2$ & $82.25$ & $-69.50$ & $34.7\pm6.2$ & $122.5\pm8.3$ & -- & AGB stars \\
    \citet{olsen02} & $79.91$ & $-69.45$ & $35.8\pm2.4$ & $145\pm4$ & -- & optical red clump \\
    \citet{nikolaev04} & $79.40$ & $-69.03$ & $30.7\pm1.1$ & $151.0\pm2.4$ & -- & Cepheids \\
    \citet{koerwer09} & $80.89$ & $-69.75$ & $23.5\pm0.4$ & $154.6\pm1.2$ & -- & NIR red clump \\
    \citet{rubele12} & $79.40$ & $-69.03$ & $26.2\pm2.0$ & $129.1\pm13.0$ & -- & fit of early VMC data \\ 
    \citet{subramanian13} & $79.91$ & $-69.45$ & $25.7\pm1.6$ & $141.5\pm4.1$ & -- & NIR red clump \\
    \citet{deb18} & $80.78$ & $-69.03$ & $25.110\pm0.365$ & $154.702\pm1.378$ & -- & multi-$\lambda$ Cepheids \\ 
    \citet{choi18a} & $82.25$ & $-69.5$ & $25.86^{+0.73}_{-1.39}$ & $149.23^{+6.43}_{-8.35}$ & -- & optical red clump
    \\ 
    \hline
    This work, all $Y\ks$ & $79.40$ & $-69.03$ & $23.99 \pm 0.33 $ & $141.40_{-0.96}^{+0.98}$       & $51.21 \pm 0.02$  \\
    This work, $Y\ks$ filtered & $79.40$ & $-69.03$ & $24.38 \pm 0.34$ & $141.19_{-0.94}^{0.96}$       & $51.15 \pm 0.02$ \\
    This work, all $J\ks$ & $79.40$ & $-69.03$ & $24.06 \pm 0.34$       & $149.52_{-0.94}^{0.93}$       & $50.58 \pm 0.02$  \\
    This work, $J\ks$ filtered & $79.40$ & $-69.03$ & $23.92 \pm 0.34$ & $149.41_{-0.93}^{+0.95}$ & $50.51 \pm 0.02$ \\
    \hline
    \end{tabular}
    \\ Notes: $^1$ $R_0$ are the heliocentric distances to the LMC centre. We omit the values assumed/derived by other authors, since they rely on very different (and sometimes outdated) distance calibrations.
    $^2$ This is the plane assumed by \citetalias{HZ09} (see Sect.~\ref{sec:sfh_hz}).
\end{table*}

\begin{figure*}
	\includegraphics[width=0.48\textwidth]{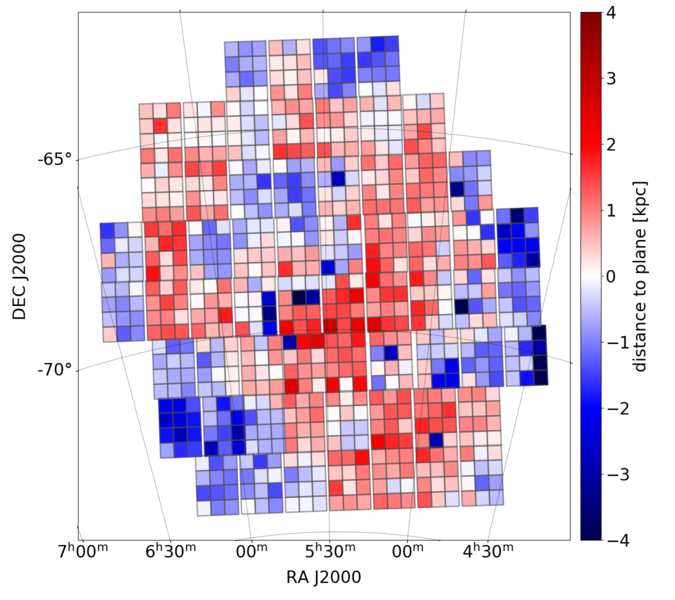}
	\includegraphics[width=0.4\textwidth]{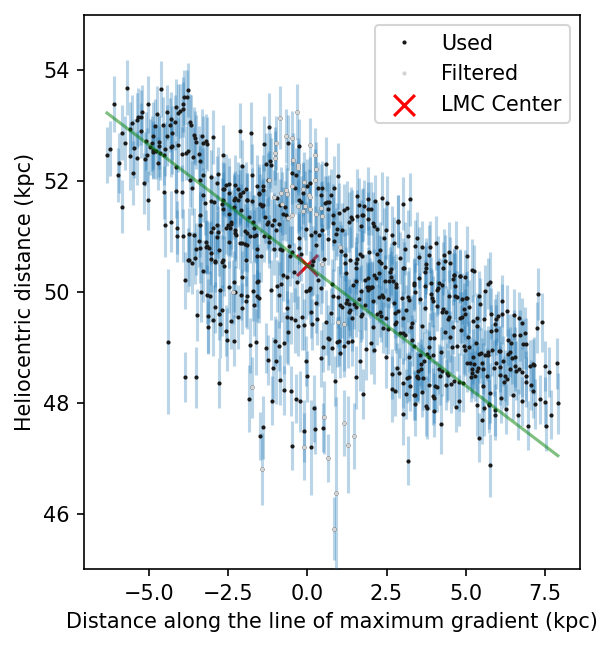}
    \caption{Left panel: residuals from the fitting of the LMC plane using the most reliable regions and $J\ks$ photometry. Right panel: heliocentric distance versus distance along the line of maximum gradient for this same fit.}.
    \label{fig:mu0_toplane}
\end{figure*}

Results are presented in Table~\ref{tab:plane}, where we also compare them with a series of similar determinations, all based on fitting a plane to the distances of indicators distributed across the LMC disc. Discussing these results in detail is beyond the scope of this paper. We just notice that the plane we fit is in better agreement with previous determinations based on the NIR magnitude of the RC \citep{koerwer09,subramanian13}, than on determinations based on Cepheids, or on optical data. This is not surprising because the $E(\ks)$ values we derive (and hence the $\mu_0$ from eq.~\ref{eq:mu0}) are strongly constrained by the \ks\ magnitude of the RC in VMC data. Our results, however, differ from determinations based \textit{only} on the RC magnitude, because (1) we consider stars in many other evolutionary stages in the fitting, and (2) we naturally take into account the intrinsic variations in the RC absolute magnitude as a function of population age and metallicity \citep[see][]{girardi16}.
 
Our best results are illustrated in Fig.~\ref{fig:mu0_toplane}. This fitted plane is expected to provide a first-order correction to the distance of all LMC stars.
 
\subsection{Spatial comparison with the \citetalias{HZ09} SFR$(t)$}
\label{sec:sfh_hz}

\begin{figure*}
	\includegraphics[width=0.33\textwidth]{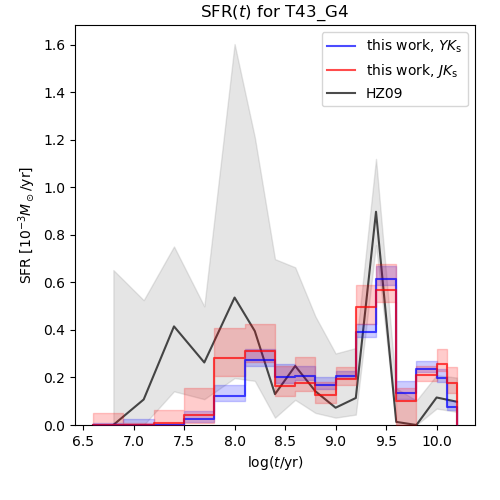}
	\includegraphics[width=0.33\textwidth]{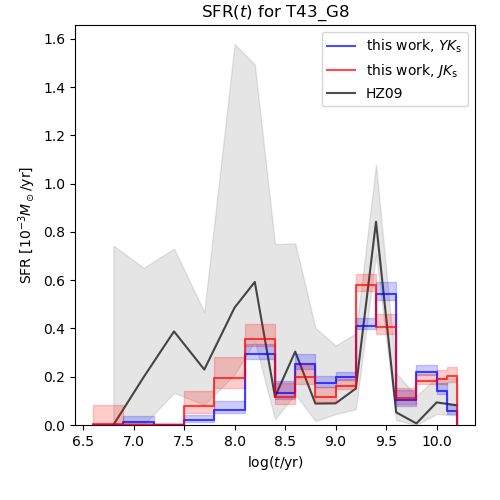}
	\includegraphics[width=0.33\textwidth]{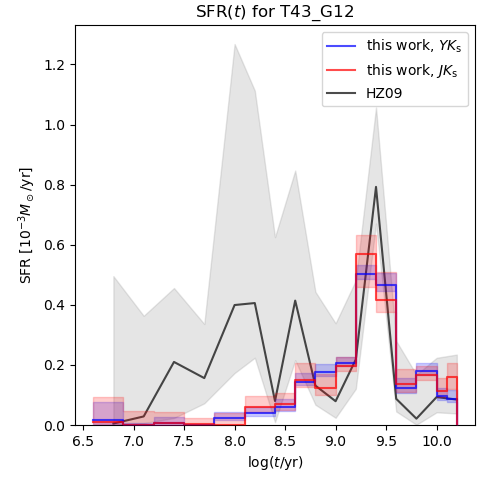}\\
	\includegraphics[width=0.33\textwidth]{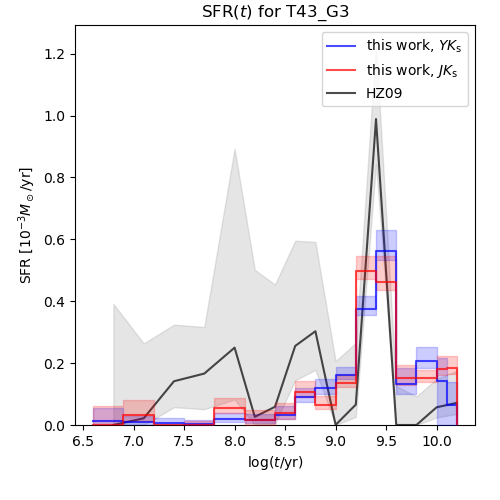}
	\includegraphics[width=0.33\textwidth]{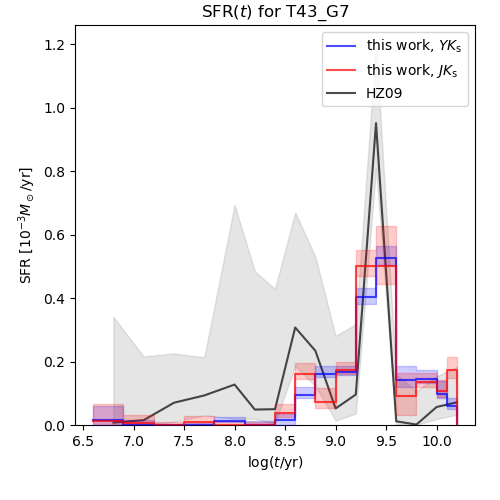}
	\includegraphics[width=0.33\textwidth]{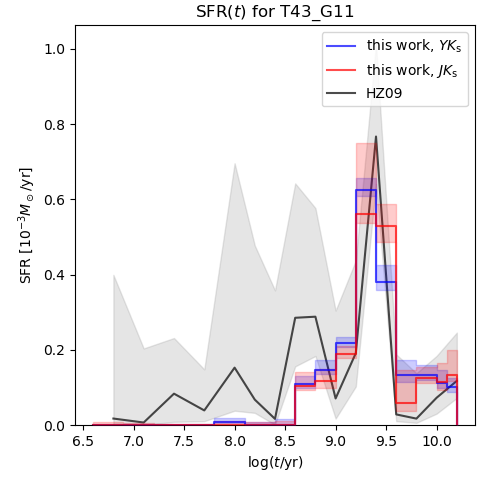}\\
	\includegraphics[width=0.33\textwidth]{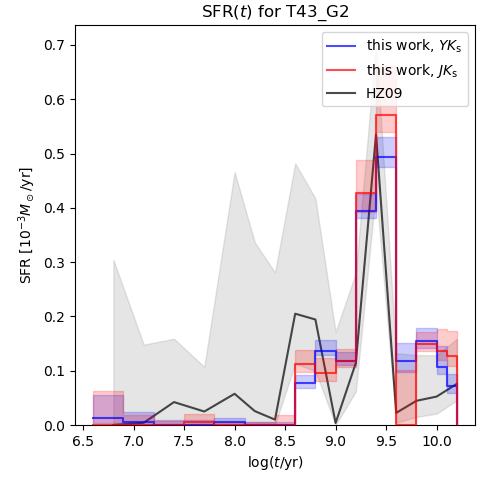}
	\includegraphics[width=0.33\textwidth]{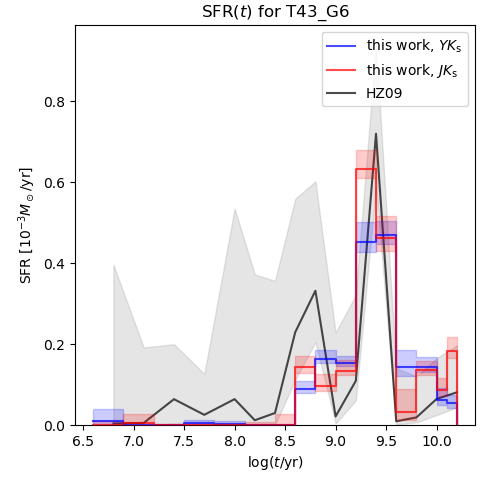}
	\includegraphics[width=0.33\textwidth]{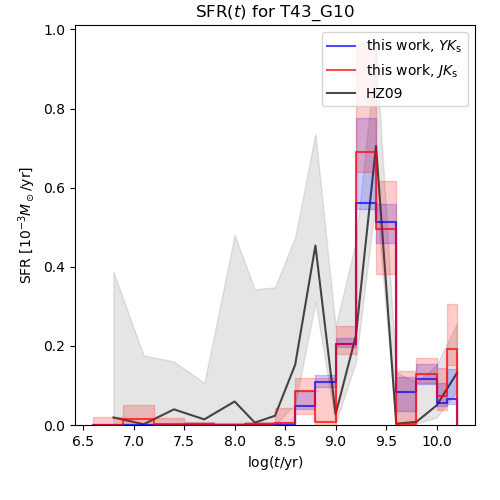}
	\\
    \caption{Comparison between our SFR$(t)$ and those of \citet{HZ09}, for subregions covering the northern 3/4 of the tile LMC 4\_3. They are plotted in the same way as they appear in the sky, with northermost subregions at the top and westernmost subregions to the right. The different curves illustrate our derived SFR$(t)$, and the \citetalias{HZ09} SFR$(t)$ resampled for the same area. The full lines correspond to the median values of the SFR$(t)$, while the shaded areas correspond to the 68\% confidence intervals.}
    \label{fig:compHZ1}
\end{figure*}

\begin{figure*}
	\includegraphics[width=0.33\textwidth]{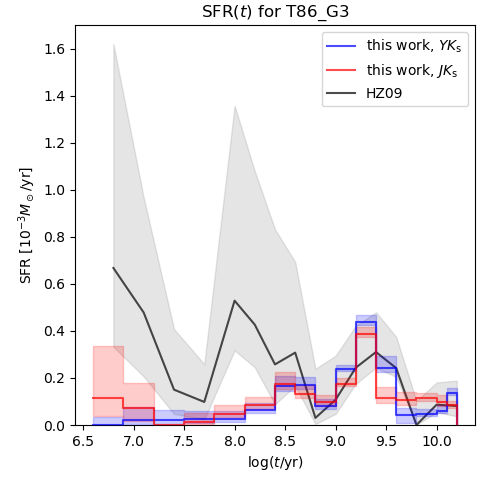}
	\includegraphics[width=0.33\textwidth]{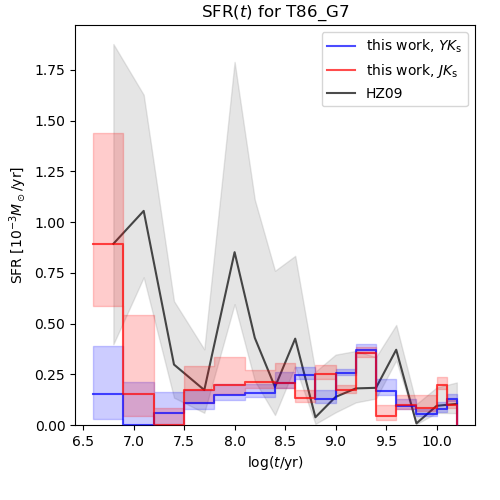}
	\includegraphics[width=0.33\textwidth]{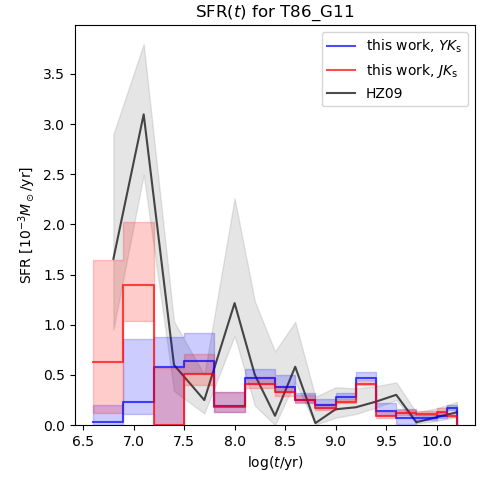}\\
	\includegraphics[width=0.33\textwidth]{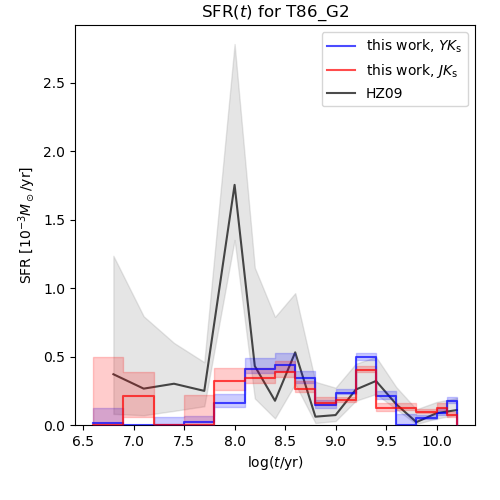}
	\includegraphics[width=0.33\textwidth]{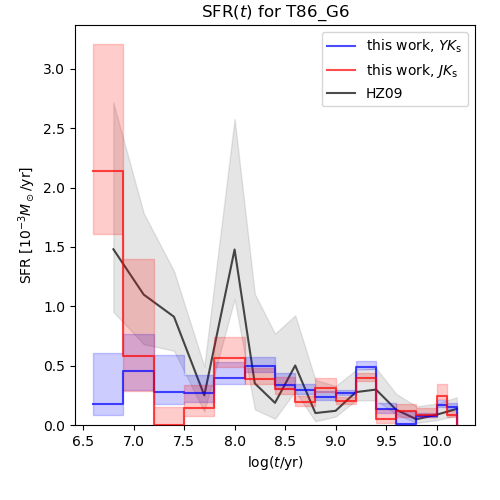}
	\includegraphics[width=0.33\textwidth]{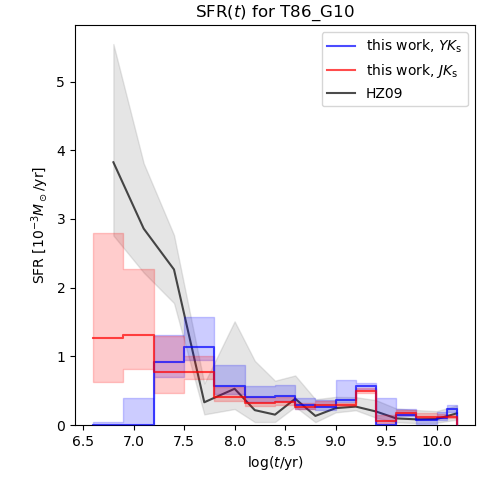}\\
	\includegraphics[width=0.33\textwidth]{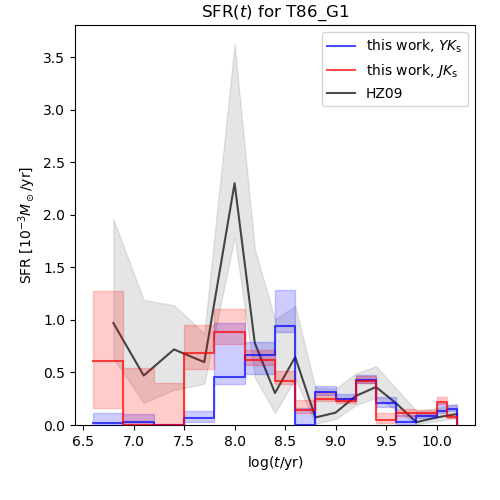}
	\includegraphics[width=0.33\textwidth]{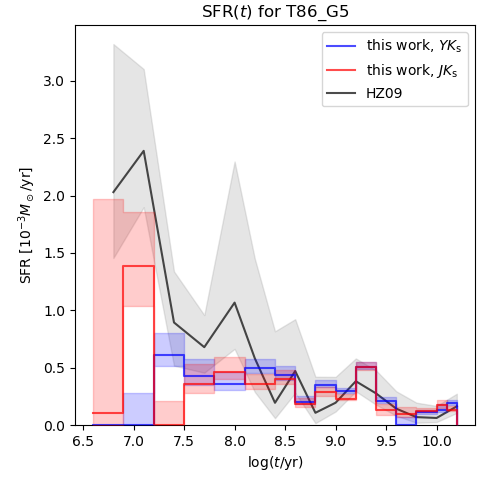}
	\includegraphics[width=0.33\textwidth]{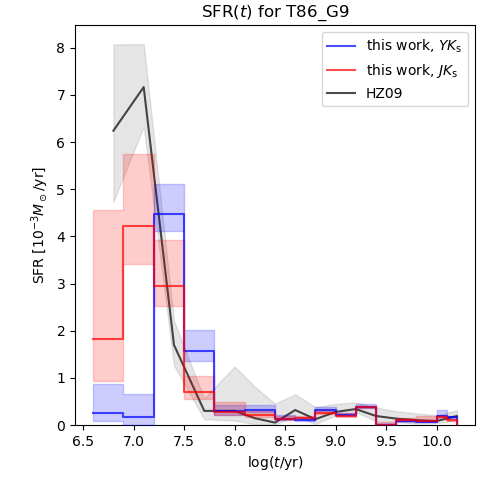}
    \caption{The same as Fig.~\ref{fig:compHZ1} but for the southern section of the tile LMC 8\_6, covering part of the star forming area known as Constellation~III.}
    \label{fig:compHZ2}
\end{figure*}

As already mentioned, the most comprehensive study of the SFH across the LMC so far is from \citetalias{HZ09}, performed for an $8.5^\circ\times7.5^\circ$ area of the optical MCPS survey. That study samples the same age range as VMC. Their survey area was divided in cells $24'\times24'$ large, or 1/4 of that size for the innermost LMC regions. The total area explored by \citetalias{HZ09} is compared to ours in Fig.~\ref{fig:compareas}.

By comparing our SFH maps of Figs.~\ref{fig:SFH_JKs_maps} and \ref{fig:SFH_YKs_maps}, with those in figure 8 of \citetalias{HZ09}, it is clear that both works present some similar features, especially at young ages. To allow a quantitative comparison with our results, we resample the SFH maps from \citetalias{HZ09} for the same areas of our subregions. It means that for every subregion, we find all the spatial cells defined by \citetalias{HZ09} that intersect it, and add the fraction of \citetalias{HZ09}'s \SFRt\ corresponding to the intersecting area.

Representative comparisons between the derived \SFRt\ are shown in Fig.~\ref{fig:compHZ1} and \ref{fig:compHZ2}, for two sections of tiles uniformly covered in both works -- namely the northernmost section of tile LMC 4\_3, and the southernmost section of tile LMC 8\_6. 
We use the same plotting approach of \citetalias{HZ09}, who simply connect their values of \logtyr\ with values of inferred \SFRt. Although their \SFRt\ values actually apply to \logtyr\ intervals $0.2$ or $0.3$~dex wide (and not to a single point in age), it is not clear exactly where these age intervals start and finish. This makes the quantitative comparison somewhat uncertain.

The results of these comparisons vary a lot, going from the ``excellent agreement'' apparent for a few subregions, to surprising differences even for the very young \SFRt, which should have been well sampled by both VMC and MCPS.
In the case of tile LMC 4\_3 (Fig.~\ref{fig:compHZ1}), the agreement appears quite good for all ages older than $1$~Gyr: there was a strong peak of \SFRt\ for ages between $2$ and $4$~Gyr ($9.3<\logtyr<9.6$), preceded by much smaller values of \SFRt\ at older ages. At younger ages, there seems to be a discrepancy at ages of $\logtyr\simeq8.6$, in which \citetalias{HZ09} suggests a peak of \SFRt, at least in the southern part of the tile, which is absent from our maps. At even younger ages, for most of the subregions we only have upper limits to the \SFRt, but these upper limits are much smaller than the upper limits found by \citetalias{HZ09}.
We identify an episode of enhanced \SFRt\ at ages $8.1<\logtyr<8.4$~yr at the top (northern) part of this tile, which has a counterpart in the $10^8$~yr peak shown by \citetalias{HZ09} maps. 
Fig.~\ref{fig:compHZ1} also illustrates that, at least for this tile, the \SFRt\ obtained from $Y\ks$ and $J\ks$ datasets are about the same.

The tile LMC 8\_6 covers a good fraction of the star-forming region known as Constellation~III. There are striking similarities with the \SFRt\ derived by \citetalias{HZ09} at all intermediate and old ages -- with the absolute values of \SFRt\ differing by less than expected from their 68\% confidence intervals (Fig.~\ref{fig:compHZ2}). At ages around $10^8$~yr, however, \citetalias{HZ09} maps tend to present stronger peaks of SFR than our maps suggest. At the even younger ages of $\logtyr<7.2$, our maps capture the presence of bursts of \SFRt\ in the same places as \citetalias{HZ09}, but with a somewhat reduced intensity, and only in the $J\ks$-derived maps.  

The quantitative comparison between our SFR$(t)$ and the \citetalias{HZ09} one is more complicated than suggested in these figures, because of a series of significant differences in the analyses. A fundamental one is in the IMFs used to model the single-burst stellar populations: a normalised \citet{kroupa01} in our case, \citet{salpeter} in the case of \citetalias{HZ09}. Different IMFs imply that the multiplicative constants that are used to convert the numbers of observed (and fitted) stars into a total initial mass of a stellar population -- and hence into a given \SFRt\ in units of $\Msun\mathrm{yr}^{-1}$ -- should have been different. Nonetheless, the final results for the \SFRt\ appear on a comparable scale in Figs.~\ref{fig:compHZ1} and \ref{fig:compHZ2}, \textit{suggesting} that \citetalias{HZ09} normalised their IMF in a way similar to us. Unfortunately, details of this IMF normalisation are not specified in their work.

There are many other differences between our derivation of the \SFRt, and those by \citetalias{HZ09}, among which: our metallicity distribution is constrained around a reference AMR, while in \citetalias{HZ09} the metallicities can simultaneously occupy four different values at every age; we fit the mean values of extinction and distance independently for every subregion, whereas \citetalias{HZ09} apply pre-defined corrections to these parameters; our studies use different generations of stellar models and methods to find the best solutions; over the Bar, \citetalias{HZ09} fix the shape of their old \SFRt, for $t>4$~Gyr, so as to replicate the results obtained by other authors using deeper HST observations, while we use \textit{only} VMC data to derive all quantities. The most notable difference, however, is that \citetalias{HZ09} use optical data, while we use the near-infrared. Optical data may provide a better colour separation for young populations, but on the other hand it is more affected by interstellar extinction and its dispersion inside the young LMC disk. Therefore, the advantage of using either one of these datasets is not obvious. It is noteworthy and reassuring that, despite all these differences, the resulting \SFRt\ do not appear dramatically different, at least not for the $t>10^8$~yr ages that contain most of the stellar mass formed in the LMC.

\subsection{The problem at very young ages}
\label{sec:sfh_veryyoungproblem}

A surprising result from this comparison is that our SFH maps from $Y\ks$ data seem to underestimate the \SFRt\ for the two youngest age bins, $6.6<\logtyr<6.9$ and $6.9<\logtyr<7.2$, compared to both the $J\ks$-derived maps and the \citetalias{HZ09} results. We verify that at these very young ages all the evolved stars happen to be located above the magnitude cut of $\ks<12$~mag. As a consequence, the only information that is actually used to constrain the \SFRt\ for these young age bins are the (very small) star counts along the upper main sequence, and especially those located at colours close to zero in the $12<\ks<14$~mag interval. For all older ages, instead, the red core helium burning stars enter decisively in the Hess diagrams being analysed. While this fact does not explain the differences between the $Y\ks$ and $J\ks$ results, they actually warn us that the \SFRt\ at $\logtyr<7.2$ might be significantly less reliable than the results for all older ages.

Very young star formation is often associated with higher extinction (i.e., larger than the one measured for older populations in the same area) and/or larger extinction dispersion. These effects are still not considered in our models, and could be contributing to these systematic errors.

\begin{figure*}
	\includegraphics[width=0.48\textwidth]{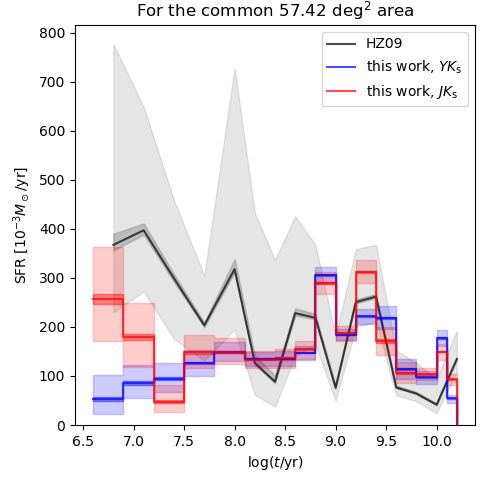}
	\includegraphics[width=0.48\textwidth]{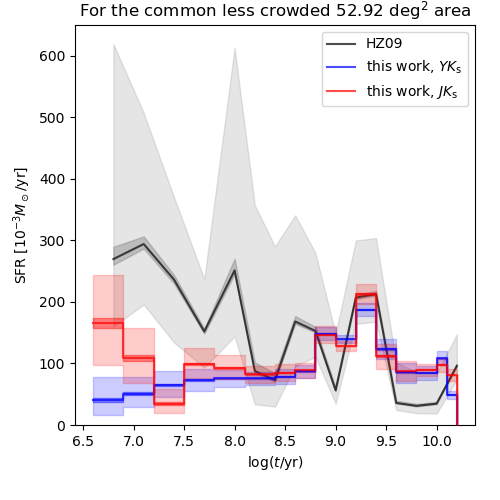}
    \caption{Comparison between our \SFRt\ and those of \citet{HZ09}, integrated over many subregions. The left panel includes the entire area in which we have results from both surveys, while in the right panels we consider only the regions with $\ks(\mathrm{75\% completeness})>18.5$~mag. In the comparison, one should better exclude the two youngest age bins, for $\logtyr<7.2$, which are considered to be significantly more uncertain than all the others.}
    \label{fig:compHZtot}
\end{figure*}
 
\subsection{Comparison with the \citetalias{HZ09} \SFRt\ for a large area}
\label{sec:sfh_hztot}

The left panel in Fig.~\ref{fig:compHZtot} presents a comparison between the two \SFRt\ integrated over a large area ($57.42$~deg$^2$) where we have results from both VMC and MCPS. Keeping in mind all the above-mentioned differences in the data and analyses, one thing is apparent: The two \SFRt\ are comparable for ages above $10^8$~yr, while at younger ages \citetalias{HZ09} tend to present larger values of \SFRt\ -- about two to three times larger than ours, although this difference is comparable with the large uncertainties that characterise results for very young populations. It is also apparent that there is a significant difference between our results for $Y\ks$ and $J\ks$ at very young ages, with the $Y\ks$ tending to miss the very young bursts that instead are detected with the $J\ks$ data, and also by \citetalias{HZ09}. The reason for this behaviour is not clear, but might be due to the stronger differential reddening that generally affects these regions with very young bursts.

At intermediate ages, the most notable difference is that our results point to a peak of \SFRt\ in the \logtyr\ interval between $8.8$ and $9.0$, which is absent from the \citetalias{HZ09} results. However, this difference is almost completely due to subregions in the LMC Bar, which are more affected by crowding, and which are analysed differently in the \citetalias{HZ09} case (with their partial use of results from HST for ages older than $4$~Gyr). Indeed, the right panel of Fig.~\ref{fig:compHZtot} presents the same comparison limited to all areas with $\ks(\mathrm{75\% completeness})>18.5$~mag, hence excluding about $4.5$~deg$^2$ over the LMC Bar. As can be seen, the peak at the $8.8$--$9.0$ \logtyr\ interval almost disappears from our \SFRt, and a better agreement with \citetalias{HZ09} is reached in this case.

\subsection{The total mass of stars ever formed}

\begin{figure*}
	\includegraphics[width=\textwidth]{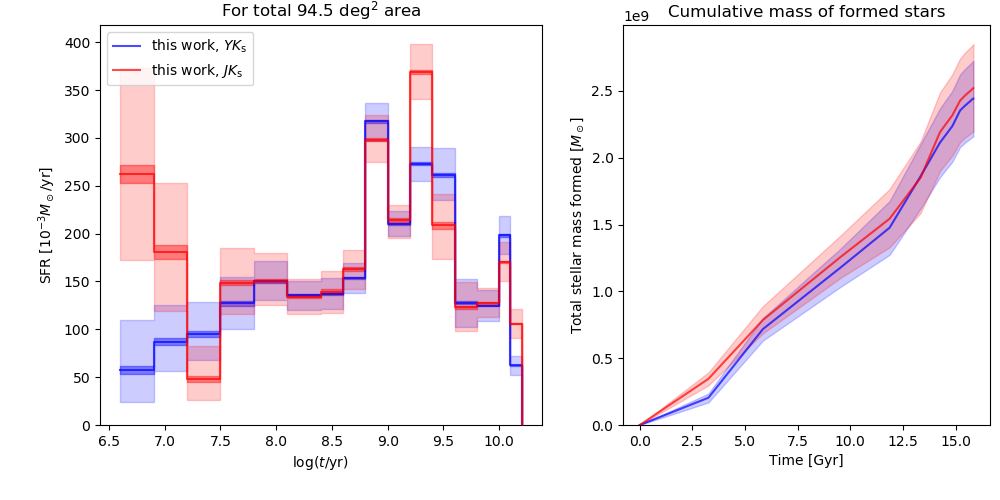}
    \caption{\textbf{Left panel:} Total \SFRt\ integrated over our total area, from both the $Y\ks$ and $J\ks$ solutions (blue and red continuous lines, respectively). The shaded areas represent the errors (or confidence regions) computed with quite different approaches: either simply adding minimum/maximum intervals for all subregions, just as if all errors were systematic (light shaded areas), or adding errors in quadrature, as if all errors were random (strong shaded areas). The \textbf{right panel} shows the cumulative mass of formed stars, obtained by integration of the \SFRt\ since the epoch of galaxy formation at $\logtyr=10.2$. In this case, the confidence regions reflect the random errors only.}
    \label{fig:SFRttot}
\end{figure*}

The left panel of Fig.~\ref{fig:SFRttot} presents the total \SFRt\ integrated over the entire area analysed in the present work. The main features in this plot were already commented in comparison with similar features derived from \citetalias{HZ09}. One aspect clearer in this figure, however, is the large uncertainties about the confidence levels in the total \SFRt. If we assume that the 756 subregions analysed provide independent solutions for their \SFRt, then errors should be combined in quadrature, with the consequence that the total \SFRt\ presents extremely narrow confidence regions. In this case, the $Y\ks$ and $J\ks$ results are generally in disagreement. If instead we consider errors as mostly systematic, and simply add the confidence intervals region by region, we obtain final errors in the total \SFRt\ which are much larger, and which generally allow for ``agreement'' between $Y\ks$ and $J\ks$ results. The correct alternative is likely somewhere in between these two cases.

The total \SFRt\ can be integrated from the epoch of LMC formation at $\logtyr=10.2$, to derive the total mass in stars formed as a function of time, which is depicted in the right panel of Fig.~\ref{fig:SFRttot}. The total mass of stars ever formed is $\sim\!2.5\times10^9$~\Msun. If we adopt just the random errors, the mass uncertainty is of about 20~\%. However, the real uncertainties are likely much larger than that. 

A total mass of $\sim\!2.5\times10^9$~\Msun\ is consistent with the $2.2\times10^9$~\Msun\ value derived by \citetalias{HZ09} (see their figure 14). It is worth remarking that a substantial fraction of this formed mass (about a third) has since been either lost to the interstellar medium or locked into compact stellar remnants. On the other hand, the most recent star formation in the LMC takes place from mass partly recycled from previous generations. Therefore, the total mass of $\sim\!2.5\times10^9$~\Msun\ gives only an order of magnitude estimate for the stellar mass that should be present now in the LMC.

\subsection{Preliminary indications for dynamical models of the Magellanic System}

As already mentioned, the present maps might be a useful resource to constrain the past history of the Magellanic System via their hydrodynamical simulations. This topic is well beyond the scope of this paper. However, looking at our results, and at the comparison with previous SFH maps from \citetalias{HZ09}, two aspects appear more relevant and worth of attention:

1) Our maps show a remarkably clear spatially-resolved \SFRt\ for all ages between a few tens of Myr and $\sim\!2$~Gyr. Previous simulations of the LMC-SMC interaction demonstrated the last strong LMC-SMC interaction occurred about 0.1-0.2 Gyr ago, possibly triggering massive star formation in the Clouds \citep[e.g.][]{yoshi03}. However, they did not predict the detailed distributions of the young stellar populations of the LMC formed by such intense interaction. The locations of the VMC tiles with \SFRt\ peaking at 0.1--0.2 Gyr can thus provide strong constraints on the LMC-SMC interaction history over the last 0.2~Gyr. We note that the more interesting age interval corresponds to our age bin $i=6$, which presents a bar and at least one clear spiral arm in  Figs.~\ref{fig:SFH_JKs_maps} and \ref{fig:SFH_YKs_maps}. 

2) The LMC bar is very well defined in the \SFRt\ maps, being slim at all ages until $\sim\!1$~Gyr, and still clearly present until $\sim\!1.6$~Gyr (Figs.~\ref{fig:SFH_JKs_maps} and \ref{fig:SFH_YKs_maps}). Although previous simulations showed the formation of a stellar bar in the LMC, in these simulations both old and new stars show strong bar-like distributions \citep{bekki05}. The observed lack of a strong bar in older populations -- if confirmed in maps less affected by crowding than ours -- would suggest that the LMC already had a dynamically hot thick disc \textit{before} the formation of the thin stellar bar \citep[see e.g.][]{bekki09}.

\section{Conclusions}
\label{sec:conclu}

In this work, we provide large-scale maps of the spatially-resolved SFH for the LMC galaxy. The main highlights of this work are:
\begin{itemize}
    \item We cover an area of $96$~deg$^2$, superseding the large area previously covered by \citetalias{HZ09} maps. 
    \item We perform a uniform analysis of the photometric data, separately for the $Y\ks$ and the $J\ks$ datasets. The VMC dataset is by itself very uniform, but for changes in observing conditions that appear as small tile-to-tile changes in the error functions and photometric zeropoints. These changes were taken into account in the analysis.
    \item Our results are generally consistent with previous works, but with some significant quantitative differences. We find a best-fitting plane for the LMC that generally agrees with those derived from near-infrared photometry of the red clump. Our \SFRt\ are similar to those of \citetalias{HZ09}, but for the trend of presenting smaller \SFRt\ values at young ages. 
    \item The periods of most intense star formation in the LMC are at intermediate ages, roughly between $0.5$ and $4$~Gyr, in which the total star formation across the entire galaxy reaches values of about $0.3$~$\Msun\mathrm{yr}^{-1}$. For the remaining epochs the total \SFRt\ is typically half of this maximum value. Two peaks of enhanced SFR$(t)$ appear in our maps, in the age intervals $8.8<\logtyr<9.0$ ($0.63$ to $1$~Gyr) and $9.2<\logtyr<9.6$ ($1.6$ to $4$~Gyr). Regions in the LMC Bar are the main responsible for the youngest of these peaks. Our results do not seem to contradict any of the general conclusions reached by \citetalias{HZ09}, but they suggest corrections to the ages of periods of enhanced SFR$(t)$, and to their quantitative values.
    \item The SFH maps are provided as a useful resource for modelling the present structure of the LMC, and its past history through chemo-dynamical evolutionary models of the LMC+SMC+MW system. Particularly interesting are the distributions of the stellar populations with ages 0.1--0.2 Gyr, which can be used to constrain the interaction histories of the LMC and the SMC, and the slim bar observed at ages younger than $\sim\!1$~Gyr, which can provide strong constrains on the dynamical properties of the LMC during/before its interaction with the SMC and the MW.
\end{itemize}

Our SFH results will be subsequently used in a series of works aimed at improving our knowledge of the Magellanic Clouds and their stars. Among these works, we will be reconsidering the recent calibration of TP-AGB evolutionary models by \citet{pastorelli20}, who used a subsample of 72 LMC subregions comprising a total area of $\sim\!9$~deg$^2$. A first comparison of the present SFH with the preliminary solution used by \citet{pastorelli20} for these subregions, reveals modest differences in the SFR$(t)$, of magnitudes similar to the random errors. More systematic changes appear in the AMR, which is now constrained to lie close to the relation indicated by spectroscopic data. Anyhow the impact of the new SFH on the TP-AGB models is still not assessed, and will be the subject of a future study. We expect that a potential larger effect may result from the fact that the LMC area useful for the TP-AGB calibration -- given by the intersection of the VMC survey area with the AGB catalog classified by \citet{boyer11} -- can now be increased by a factor of about 6. Such an increase in area results in a substantial improvement in the statistics and will also make it possible to calibrate the TP-AGB models in subregions characterized by  very different mean ages, and hence different mean progenitor masses.

The analyses of the present dataset represented a major effort in terms of processing of VMC data, and its fitting via a long process involving many different components derived from theoretical models. Simplifications were necessary to deal with the entire dataset without running out of computer storage and CPU time. Some of these simplifications will be progressively abandoned in the near future, as we add more VMC tiles to our database, and as we reanalyse present tiles in the search of better solutions. In particular, we are currently planning to perform more ASTs so as to include fainter sections of the Hess diagrams, to include 2MASS data in $J\ks$ passbands to better probe the very young SFR$(t)$, to split the data in smaller subregions so as to better deal with variable extinction in central parts of the LMC, and to separate all known star clusters from the field. Moreover, we can also implement improvements in the fitting process itself, starting from the initial solutions derived for this paper: It is somewhat evident that we can aim at a better description of the spatial distribution of stars and dust inside every subregion, also adopting constraints from neighbouring subregions. On the other hand, the new accuracy standards set by Gaia parallaxes \citep{gaiaDR2-HRD,gaiaEDR3} and by asteroseismology \citep{khan18,ted21}, and the realisation of the importance of fast-rotating stars \citep{costa19}, are prompting further revision of our stellar models, which will have its implications for evolutionary models of LMC stars. For all these reasons, we plan to reanalyse the entire VMC dataset of 104.8~deg$^2$ across the LMC in less than two years time.

\section*{Data availability}
\label{sec:dataavail}
The image data used in this paper are available in the VISTA Science Archive (VSA), at \url{http://horus.roe.ac.uk/vsa}. The PSF photometry and ASTs will be soon shared by ESO in the regular VMC data releases (\url{http://www.eso.org/rm/publicAccess#/dataReleases}).

\section*{Acknowledgements}
Many of us acknowledge the support from the ERC Consolidator Grant funding scheme (project STARKEY, grant agreement n. 615604).  M-RC and CPMB acknowledge support the European Research Council (ERC) under the European Unions' Horizon 2020 research and innovation programme (grant agreement no. 682115). 
This work is based on the observations collected at the European Organisation for Astronomical Research in the Southern Hemisphere under ESO programme 179.B-2003. We thank the CASU and the WFAU for providing calibrated data products under the support of the Science and Technology Facility Council (STFC) in the UK.



\bibliographystyle{mnras}
\bibliography{references}

\bsp	
\label{lastpage}

\clearpage\newpage


\appendix

\section{The impact of different assumptions}
\label{sec:apperrors}

Here we discuss how our results would change by altering a couple of preceding assumptions.

First, we did not adopt extinction maps nor any pre-defined model for the distances. Both parameters can be fixed, a posteriori, so as to force our solutions to be located exactly on the mean LMC plane defined in Fig.~\ref{sec:geometry}. This choice would also replicate the assumptions of known extinction and distance adopted -- although using different extinction maps and LMC plane -- by \citetalias{HZ09}. Second, in precedence we allowed the model AMR relation to vary from subregion to subregion. Since no significant large-scale metallicity trend was derived from our fits, we can assume that all subregions follow the same AMR. To do that, we simply use the mean values of metallicity shifts mentioned in Sect.~\ref{sec:metallicity}, with different values for the $J\ks$ and $Y\ks$ datasets.

\begin{figure*}
	\includegraphics[trim=0 0.8cm 0 0,clip,width=\textwidth]{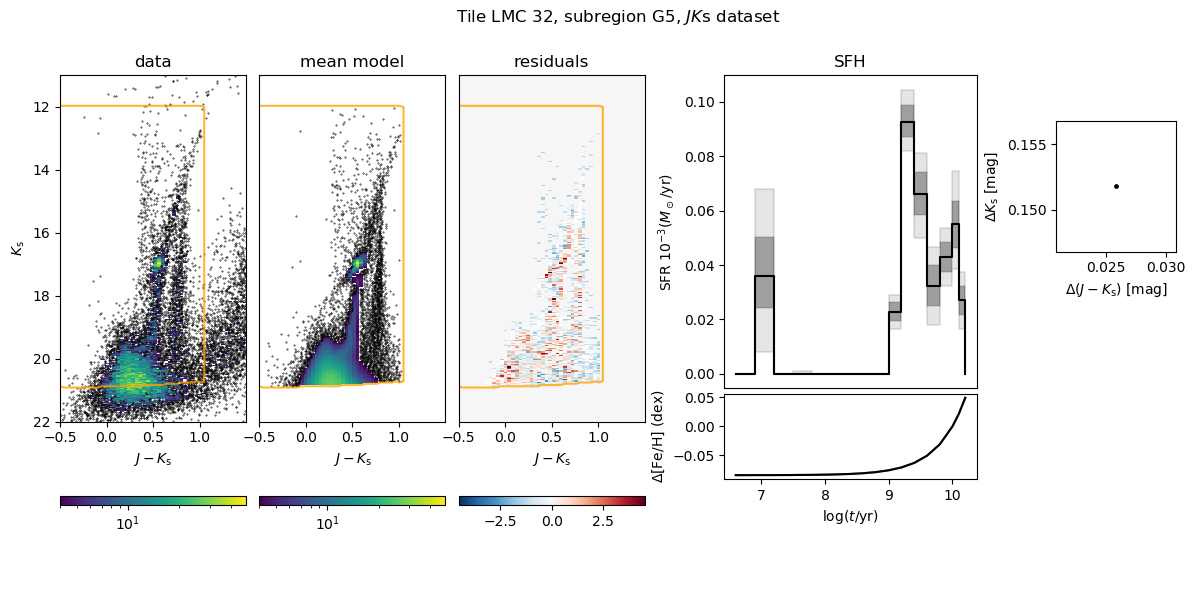}
	\includegraphics[trim=0 0.8cm 0 0,clip,width=\textwidth]{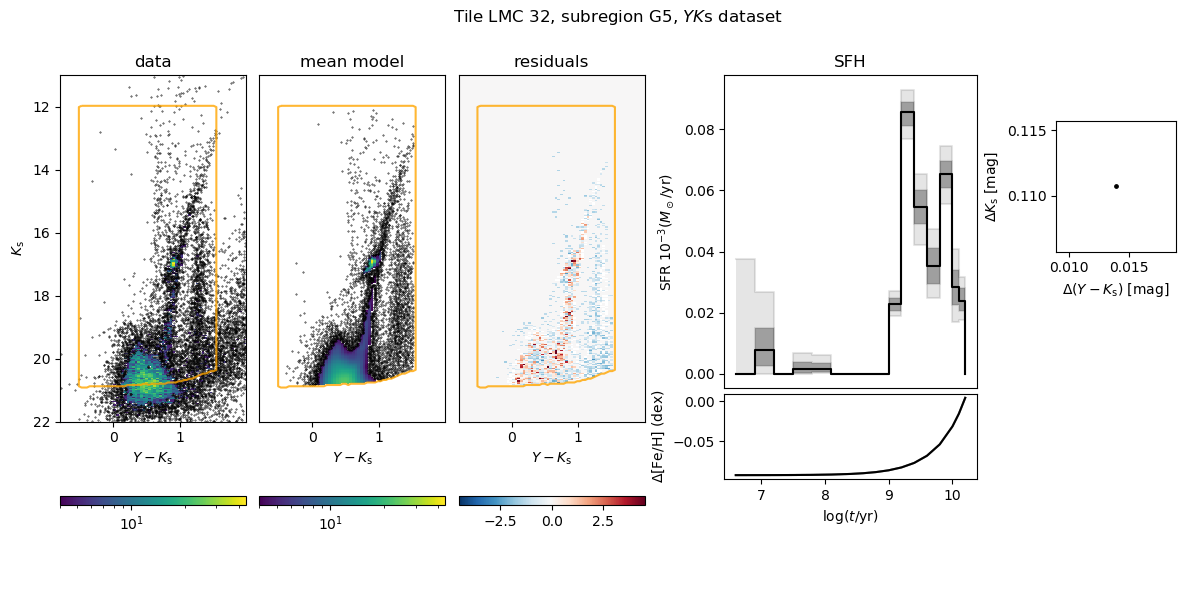}
    \caption{The same as Fig.~\ref{fig:fitexample-YKs}, but adopting fixed values of colour--magnitude and metallicty shifts (see text).}
    \label{fig:SFH_YKsfix}
\end{figure*}

\begin{figure*}
	\includegraphics[trim=0 0.8cm 0 0,clip,width=\textwidth]{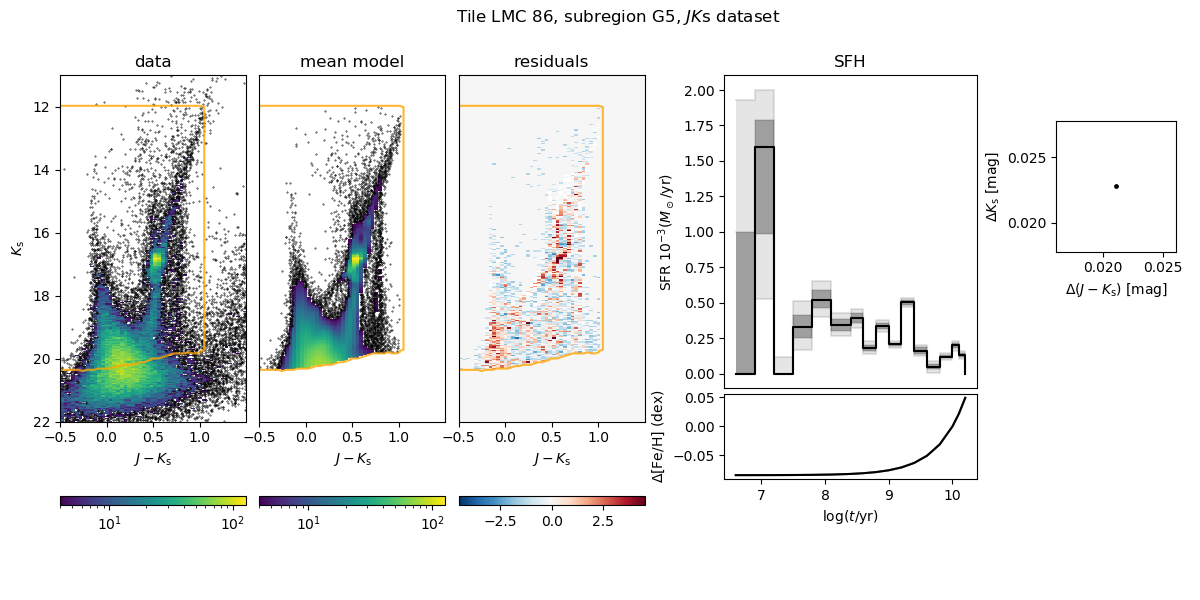}
	\includegraphics[trim=0 0.8cm 0 0,clip,width=\textwidth]{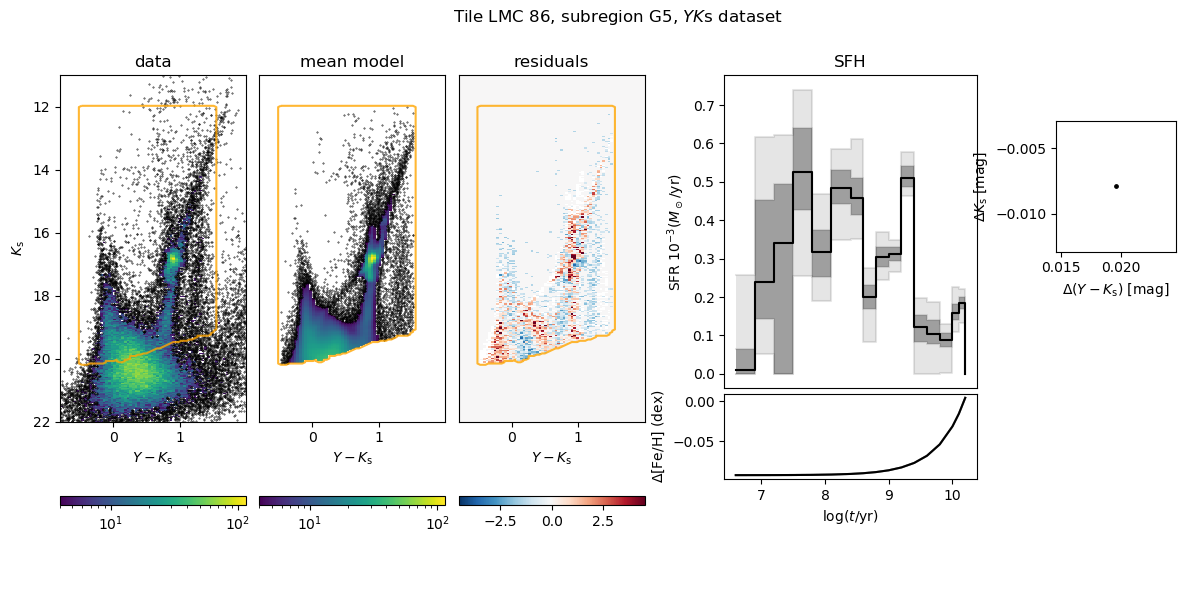}
    \caption{The same as Fig.~\ref{fig:fitexample-JKs}, but adopting fixed values of colour--magnitude and metallicty shifts (see text).}
    \label{fig:SFH_JKsfix}
\end{figure*}

Figures~\ref{fig:SFH_YKsfix} and \ref{fig:SFH_JKsfix} show the solutions resulting when these assumptions are adopted for subregions T32\_G5 and T86\_G5. They can be compared to Fig.~\ref{fig:fitexample-YKs} and \ref{fig:fitexample-JKs}, respectively. In both cases, the new \SFRt\ look very similar to those derived in precedence -- although they present slightly increased values of $-\lnL$. In general, the new median \SFRt\ lies inside the confidence intervals of the previous solutions.

\begin{figure*}
	\includegraphics[trim=0 0.8cm 0 0,clip,width=\textwidth]{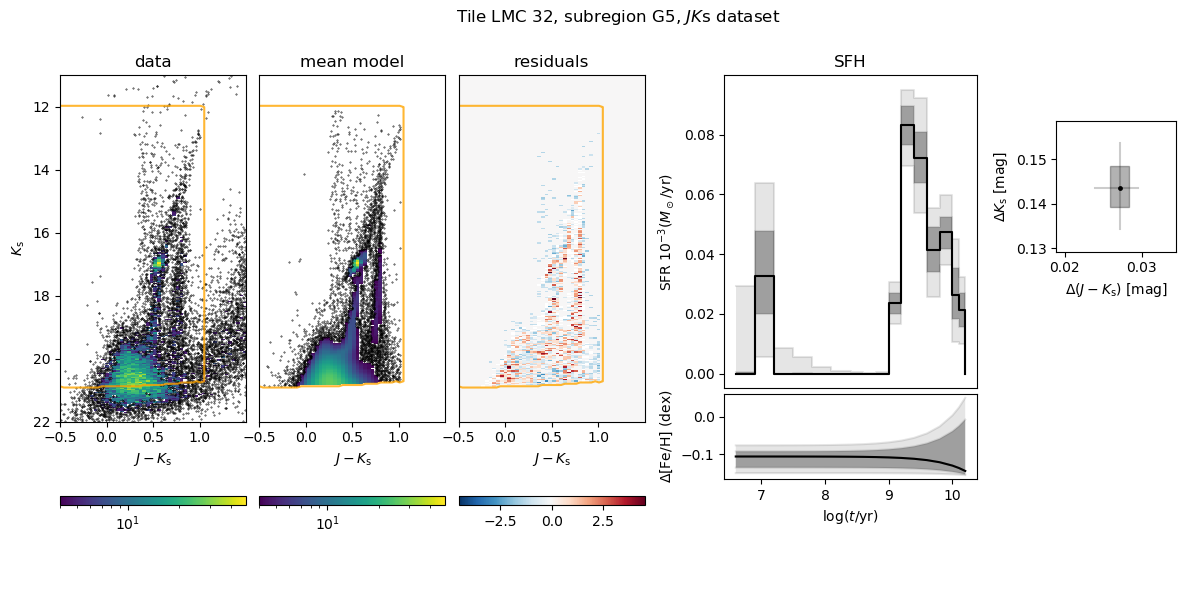}
	\includegraphics[trim=0 0.8cm 0 0,clip,width=\textwidth]{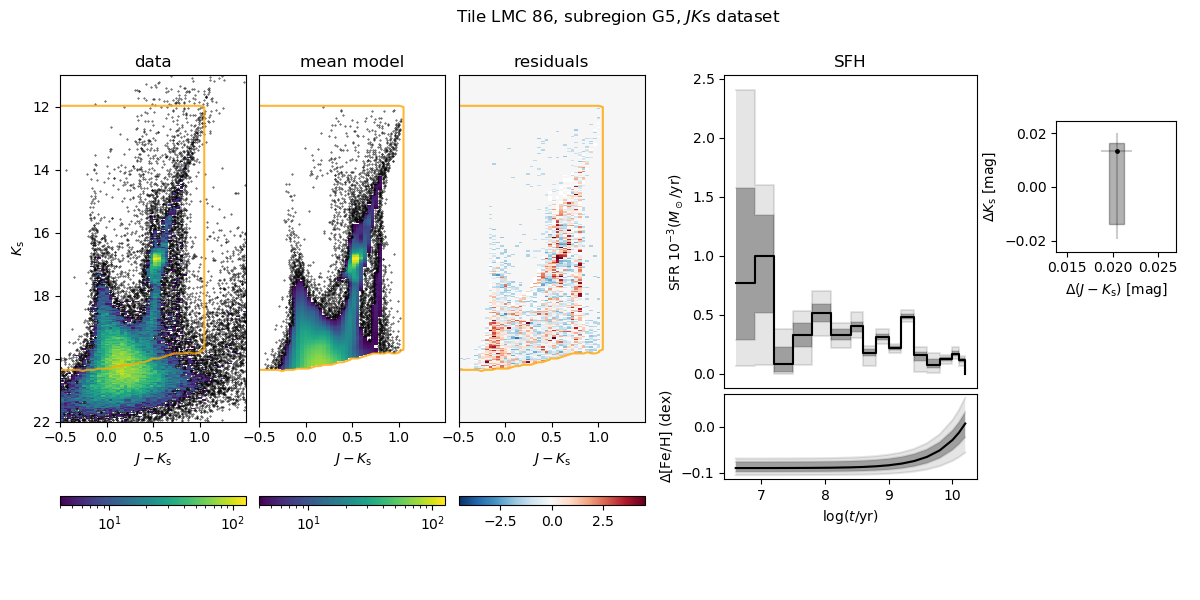}
    \caption{The same as Figs.~\ref{fig:fitexample-YKs} and \ref{fig:fitexample-JKs}, but adopting a MW foreground enhanced by a factor of 1.3 (see text). }
    \label{fig:SFH_enhancedMW}
\end{figure*}

Second, in absence of fields that could be used as pure foreground+background, we defined a $\mathbf{PM}_0$ component built from TRILEGAL models (see Sect.~\ref{sec:par2tot}). Parameters in these models are calibrated on star counts provided by multi-band wide-area surveys plus a few very-deep fields. But away from the calibration fields and their magnitude range, it is usual to find localised discrepancies of $\sim\!20$ per cent between predicted and modelled star counts. In order to evaluate the impact of this uncertainty on the SFHs, we run the CMD fitting adopting enhanced/reduced foregrounds. Fig.~\ref{fig:SFH_enhancedMW} shows the effect of enhancing the entire MW foreground by a multiplicative factor of 1.3, for the $J\ks$ datasets of subregions T32\_G5 and T86\_G5. In these plots, one can appreciate the higher residuals that appear along the reddest vertical feature, indicative of models with an overestimated MW foreground. Comparison with Figs.~\ref{fig:fitexample-YKs} and \ref{fig:fitexample-JKs} indicates that the resulting \SFRt\ solutions are essentially the same as in the models with the standard MW foreground. We also find very similar solutions when we adopt a reduced foreground (tested with a multiplicative factor of 0.7), and for the $Y\ks$ dataset.

\begin{figure*}
	\includegraphics[trim=0 0.8cm 0 0,clip,width=\textwidth]{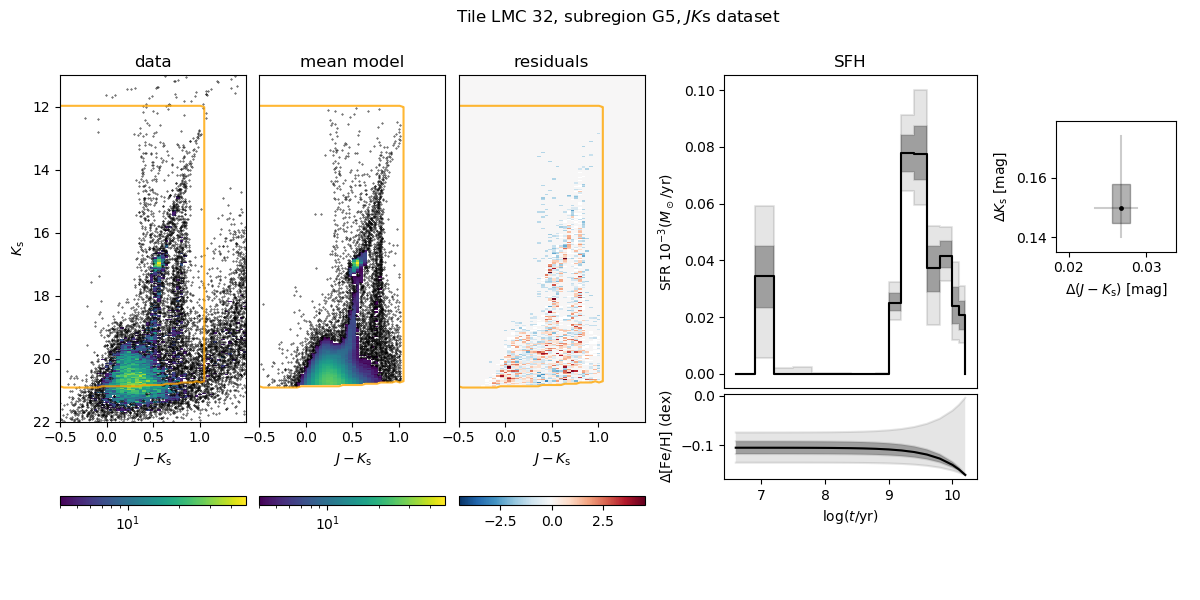}
    \caption{The same as Figs.~\ref{fig:fitexample-YKs} and \ref{fig:fitexample-JKs}, but adding a partial model for the background galaxies. }
    \label{fig:SFH_withgal}
\end{figure*}

Third, we have ignored the possible role of background galaxies. Most of them are excluded from our CMD area, especially the wide distribution of points at the bottom-right corner of our CMDs, indicatively at $\ks\ga16$~mag, $\yks\ga1.5$~mag and $\jks\ga1$~mag in Fig.~\ref{fig:cmdobs}. The problem is to determine how many galaxies are not eliminated by our cuts. Estimation of this number was recently made possible by Bell et al. (in preparation), which follows from the work described in \citet{bell19,bell20} for the SMC. They use a combination of the PSF photometry from VMC and the fitting of the optical+infrared spectral energy distribution derived from SMASH+VMC photometry, to identify and characterise galaxies in the tile LMC 3\_4. From a total of over 42,000 galaxies identified in this tile, just $\sim\!6750$ are blue and bright enough to enter in the CMD region we consider for the SFH analyses. They represent just 3 per cent of the ``stellar sources'' we have considered for this tile. This contamination is not only modest, but it also occurs at a colour-magnitude location that little overlaps with our stellar PMs. To better estimate the effect of these neglected galaxies in our results, we run a few CMD fittings where we add, to the $\mathbf{PM}_0$ model that represents the MW foreground, 1/12 of the galaxies detected in the tile LMC 3\_4. The results are in Fig.~\ref{fig:SFH_withgal}, only for the T32\_G5 case and with the $J\ks$ database. As can be appreciated, the \SFRt\ changes little in this case, compared to the previous solution (Fig.~\ref{fig:fitexample-YKs}). These changes become even smaller in subregions of higher stellar density.

From these tests, we conclude that our results are robust against reasonable changes in the definition of the MW foreground and galaxy background.

\section{Additional plots}
\label{sec:app}

\begin{figure*}
	\includegraphics[width=\textwidth]{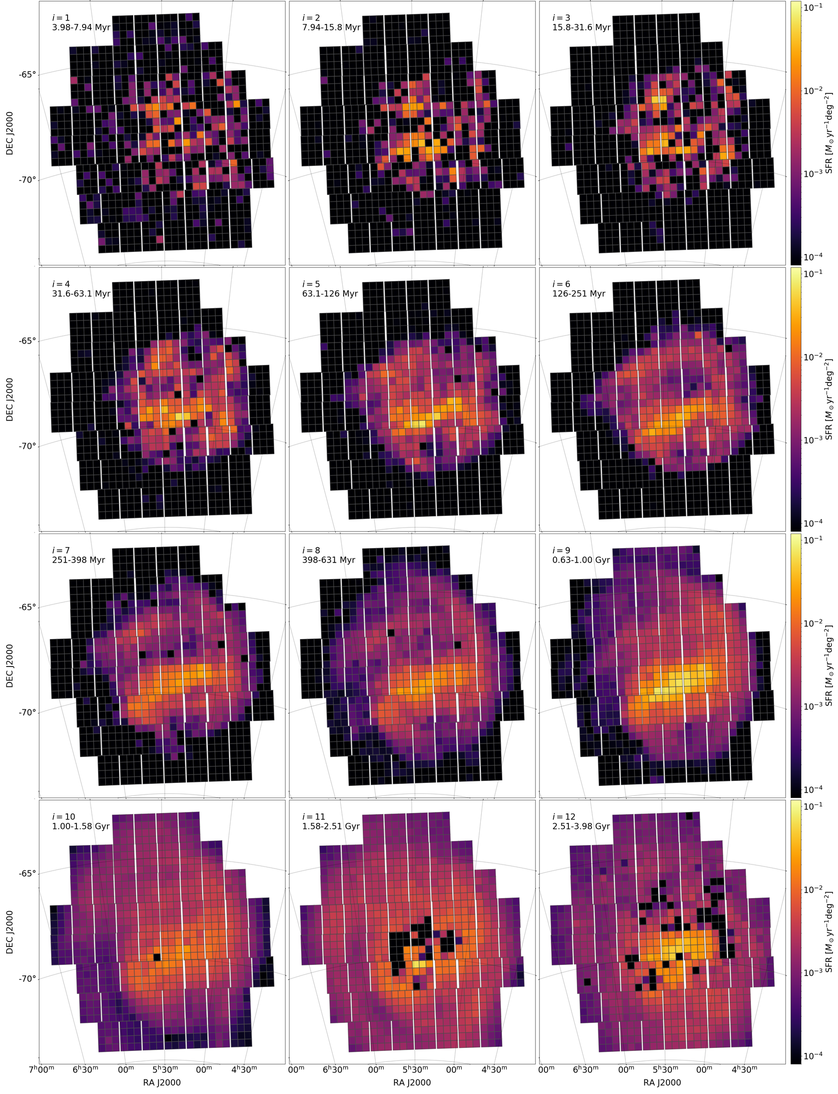}
    \caption{Same as Fig.~\ref{fig:SFH_JKs_maps} but from $Y\ks$.}
    \label{fig:SFH_YKs_maps}
\end{figure*}

\begin{figure*}
	\includegraphics[width=\textwidth]{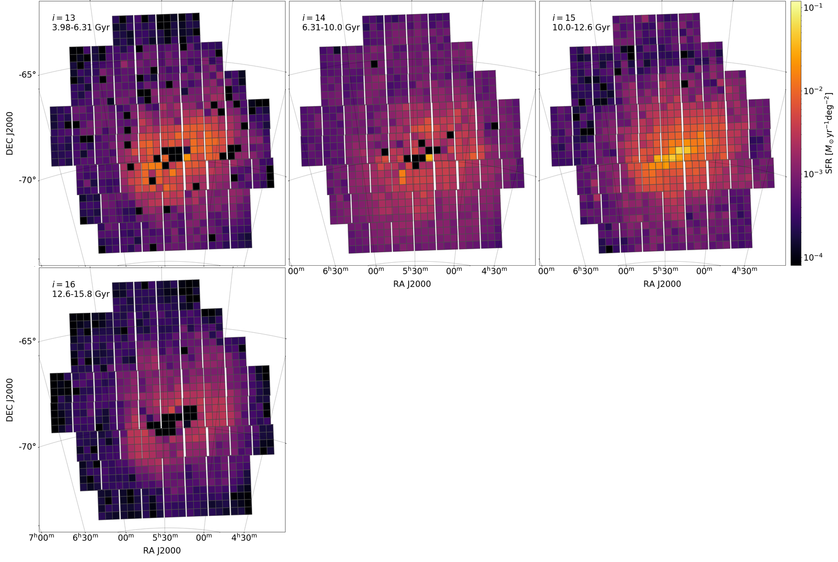}
    \contcaption{}
\end{figure*}

\begin{figure}
	\includegraphics[width=\columnwidth]{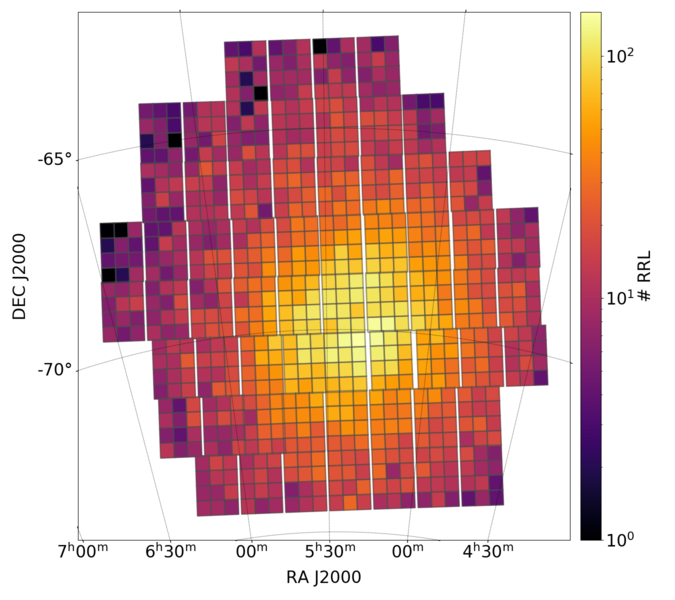}
	\includegraphics[width=\columnwidth]{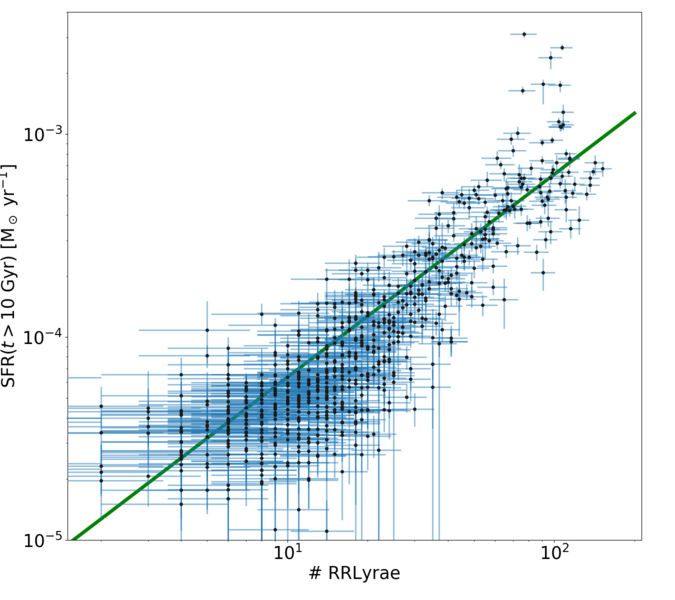}
    \caption{Top panel: Map of the known RR Lyrae per subregion, from the catalogue by \citet{cusano21}. Bottom panel: similarly to Fig.~\ref{fig:rrl_maps}, a comparison between the \SFRt\ at very old ages as derived from the $Y\ks$ data, with the number of RR Lyrae. The slope of the mean relation (green line) is of $1.58\times10^{5}$ RR Lyrae per unit $\Msun\mathrm{yr}^{-1}$. }
    \label{fig:rrl_addmaps}
\end{figure}

\begin{figure}
	\includegraphics[width=\columnwidth]{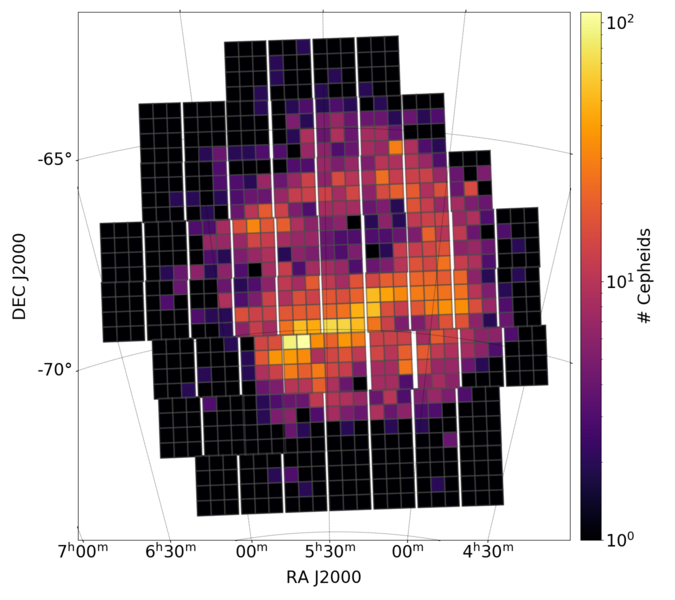}
	\includegraphics[width=\columnwidth]{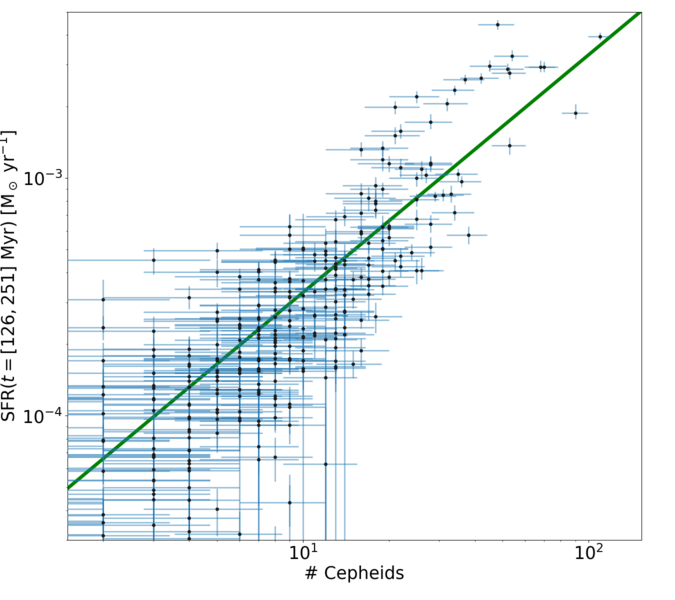}
    \caption{Top panel: Map of the known Cepheids per subregion, from the catalogue by Ripepi et al. (in preparation). Bottom panel: similarly to Fig.~\ref{fig:cepheids_maps}, a comparison between the \SFRt\ at ages between $126$ and $251$~Myr, as derived from the $Y\ks$ data, with the number of Cepheids.  The slope of the mean relation (green line) is of $3.03\times10^{4}$ Cepheids per unit $\Msun\mathrm{yr}^{-1}$.}
    \label{fig:cepheids_addmaps}
\end{figure}

This section contains plots that are not essential in the main body of this paper, but which make an integral part of our results and are as good as the plots already shown:
\begin{itemize}
    \item Figure~\ref{fig:SFH_YKs_maps} shows the maps of the \SFRt\ derived from the $Y\ks$ dataset, similarly to the maps derived from $J\ks$ already shown in Fig.~\ref{fig:SFH_JKs_maps}.
    \item Figure~\ref{fig:rrl_addmaps} presents the spatial distribution of RR Lyrae and their correlation with the old SFR derived from the $Y\ks$ data, complementing the comparison shown in Fig.~\ref{fig:rrl_maps} for $J\ks$ data. It is evident that the spatial scale of the distribution of RR Lyrae is very similar to the old \SFRt\ depicted in the two last panels of Fig.~\ref{fig:SFH_YKs_maps}.
    \item Figure~\ref{fig:cepheids_addmaps} presents the spatial distribution of Cepheids and their correlation with the \SFRt\ for the age bin between $126$ and $251$~Myr from the $Y\ks$ data, complementing the comparison shown in Fig.~\ref{fig:cepheids_maps} for $J\ks$ data. It is evident that the the distribution of the Cepheids is similar to the \SFRt\ map of $i=6$ in Fig.~\ref{fig:SFH_YKs_maps}.
\end{itemize}
The equivalent of Figs.~\ref{fig:fitexample-YKs}, \ref{fig:fitexample-JKs} and \ref{fig:fitexample-highextinction}, for all subregions, are available as single compressed file upon request to the main author.

\section{Online tables}
\label{sec:catalog}
 
Table~\ref{tabellone} presents most of the results discussed in this paper. For all subregions, they include the coordinates of the corners, the \SFRt\ in 16 age bins, the true distance moduli and extinction values, the metallicity shifts, together with their 68 per cent confidence intervals. This particular table is available in machine-readable form in the journal website, and refers to the median values obtained from the $J\ks$ data. Additional tables, including the results from the $Y\ks$ data and the limits of 68\% confidence intervals, will be available at the VizieR database of astronomical catalogues at the Centre de Donn\'ees astronomiques de Strasbourg (CDS).

\setlength{\tabcolsep}{2pt}
\begin{table*}
\caption{Sample table summarising the results of the SFH analysis.}
\label{tabellone}
\begin{scriptsize}
\begin{tabular}{cc|ccccc|cc|ccccc|cc|cc|c}
\hline\hline
tile & subr. & 
RA$_1$ & DEC$_1$ & $\cdots$ & RA$_4$ & DEC$_4$ & Area & Number &
$a_1$ & $a_2$ & $a_3$ & $\cdots$ & $a_{16}$ &
$\mu_0$ & $A_V$ & 
$\Delta\feh_1$ & $\Delta\feh_2$ &
$-\lnL$ 
\\
LMC  &       & \degr & \degr & $\cdots$ & \degr & \degr & deg$^2$ & stars & 
      $\Msun/\mathrm{yr}$ & $\Msun/\mathrm{yr}$ & $\Msun/\mathrm{yr}$ & $\cdots$ & $\Msun/\mathrm{yr}$ &  
      mag & mag & dex & dex & \\
     &       & (1) & (1) & (1) & (1) & (1) & (2) & (3) & (4) & (4) & (4) & (4) & (4) & (5) & (5) & (6) & (6) & (7) \\
\hline
32 & 1 & 69.5447 & -73.6400 & $\cdots$ & 70.7203 & -73.7160 & 0.125 & 24784 & 2.60e-08 & 4.86e-08 & 5.80e-09 & $\cdots$ & 2.92e-05 & 18.637 & 0.089 & -0.138 & 0.155 & 2.79e+03 \\
32 & 2 & 69.8442 & -73.2813 & $\cdots$ & 70.9964 & -73.3556 & 0.125 & 28794 & 9.93e-20 & 1.02e-14 & 7.15e-17 & $\cdots$ & 1.96e-05 & 18.608 & 0.322 & -0.147 & -0.073 & 2.63e+03 \\
32 & 3 & 70.1314 & -72.9221 & $\cdots$ & 71.2612 & -72.9949 & 0.125 & 33210 & 3.96e-05 & 0.00e+00 & 0.00e+00 & $\cdots$ & 2.22e-17 & 18.598 & 0.362 & -0.126 & -0.159 & 2.99e+03 \\
32 & 4 & 70.4072 & -72.5625 & $\cdots$ & 71.5153 & -72.6339 & 0.125 & 38792 & 6.28e-06 & 2.75e-07 & 4.62e-08 & $\cdots$ & 4.76e-05 & 18.637 & 0.129 & -0.068 & -0.029 & 2.82e+03 \\
32 & 5 & 68.3801 & -73.5576 & $\cdots$ & 69.5447 & -73.6400 & 0.125 & 20474 & 1.40e-07 & 2.12e-05 & 4.16e-08 & $\cdots$ & 2.24e-05 & 18.603 & 0.353 & -0.118 & 0.014 & 2.54e+03 \\
32 & 6 & 68.7023 & -73.2006 & $\cdots$ & 69.8442 & -73.2813 & 0.125 & 24760 & 1.06e-05 & 3.76e-06 & 9.95e-07 & $\cdots$ & 5.96e-06 & 18.561 & 0.393 & -0.100 & 0.025 & 2.89e+03 \\
32 & 7 & 69.0114 & -72.8430 & $\cdots$ & 70.1314 & -72.9221 & 0.125 & 27073 & 1.96e-26 & 0.00e+00 & 3.51e-06 & $\cdots$ & 2.01e-05 & 18.586 & 0.393 & -0.158 & 0.153 & 3.01e+03 \\
32 & 8 & 69.3083 & -72.4850 & $\cdots$ & 70.4072 & -72.5625 & 0.125 & 32839 & 4.16e-07 & 1.11e-07 & 3.93e-09 & $\cdots$ & 3.11e-05 & 18.597 & 0.379 & -0.130 & -0.088 & 2.73e+03 \\
32 & 9 & 67.2272 & -73.4688 & $\cdots$ & 68.3801 & -73.5576 & 0.088 & 12226 & 1.91e-12 & 3.10e-08 & 5.33e-12 & $\cdots$ & 6.54e-06 & 18.599 & 0.383 & -0.126 & -0.128 & 2.28e+03 \\
32 & 10 & 67.5714 & -73.1136 & $\cdots$ & 68.7023 & -73.2006 & 0.125 & 20432 & 1.04e-41 & 5.39e-06 & 2.03e-44 & $\cdots$ & 1.03e-05 & 18.584 & 0.494 & -0.157 & 0.115 & 2.98e+03 \\
32 & 11 & 67.9018 & -72.7578 & $\cdots$ & 69.0114 & -72.8430 & 0.125 & 23767 & 5.83e-09 & 2.74e-06 & 2.52e-09 & $\cdots$ & 1.10e-06 & 18.599 & 0.400 & -0.131 & 0.012 & 2.83e+03 \\
32 & 12 & 68.2191 & -72.4016 & $\cdots$ & 69.3083 & -72.4850 & 0.125 & 26723 & 2.24e-05 & 0.00e+00 & 0.00e+00 & $\cdots$ & 2.29e-05 & 18.618 & 0.281 & -0.118 & -0.160 & 2.69e+03 \\
33 & 1 & 73.3617 & -73.8620 & $\cdots$ & 74.5671 & -73.9171 & 0.125 & 38848 & 0.00e+00 & 0.00e+00 & 0.00e+00 & $\cdots$ & 4.66e-05 & 18.649 & -0.029 & 0.160 & -0.140 & 2.75e+03 \\
33 & 2 & 73.5857 & -73.4986 & $\cdots$ & 74.7659 & -73.5526 & 0.125 & 44581 & 2.08e-05 & 1.70e-08 & 1.38e-10 & $\cdots$ & 4.96e-05 & 18.628 & 0.142 & 0.124 & -0.158 & 3.16e+03 \\
33 & 3 & 73.8003 & -73.1351 & $\cdots$ & 74.9563 & -73.1879 & 0.125 & 51483 & 6.12e-12 & 3.27e-12 & 6.71e-08 & $\cdots$ & 7.47e-05 & 18.606 & 0.308 & -0.049 & -0.136 & 3.10e+03 \\
33 & 4 & 74.0061 & -72.7713 & $\cdots$ & 75.1389 & -72.8230 & 0.125 & 54214 & 5.70e-06 & 3.09e-22 & 5.23e-19 & $\cdots$ & 8.87e-05 & 18.648 & -0.022 & 0.160 & -0.156 & 3.07e+03 \\
33 & 5 & 72.1647 & -73.8001 & $\cdots$ & 73.3617 & -73.8620 & 0.125 & 37867 & 9.40e-06 & 2.71e-10 & 4.77e-07 & $\cdots$ & 7.20e-05 & 18.616 & 0.245 & -0.119 & -0.116 & 2.86e+03 \\
33 & 6 & 72.4134 & -73.4381 & $\cdots$ & 73.5857 & -73.4986 & 0.125 & 43343 & 8.17e-06 & 1.21e-10 & 1.77e-06 & $\cdots$ & 4.31e-05 & 18.604 & 0.301 & -0.005 & -0.152 & 3.22e+03 \\
33 & 7 & 72.6517 & -73.0758 & $\cdots$ & 73.8003 & -73.1351 & 0.125 & 49663 & 2.06e-11 & 7.10e-11 & 2.34e-09 & $\cdots$ & 4.88e-05 & 18.610 & 0.314 & -0.076 & -0.139 & 3.24e+03 \\
33 & 8 & 72.8803 & -72.7133 & $\cdots$ & 74.0061 & -72.7713 & 0.125 & 57213 & 2.87e-05 & 1.10e-06 & 1.28e-07 & $\cdots$ & 9.72e-05 & 18.630 & 0.139 & 0.062 & -0.142 & 3.16e+03 \\
33 & 9 & 70.9771 & -73.7317 & $\cdots$ & 72.1647 & -73.8001 & 0.088 & 22940 & 3.27e-22 & 9.65e-30 & 3.71e-06 & $\cdots$ & 3.33e-05 & 18.593 & 0.238 & -0.105 & -0.059 & 2.56e+03 \\
33 & 10 & 71.2499 & -73.3711 & $\cdots$ & 72.4134 & -73.4381 & 0.125 & 38409 & 9.17e-09 & 2.26e-06 & 1.97e-06 & $\cdots$ & 4.19e-05 & 18.619 & 0.165 & 0.069 & -0.061 & 3.00e+03 \\
33 & 11 & 71.5114 & -73.0102 & $\cdots$ & 72.6517 & -73.0758 & 0.125 & 42768 & 6.90e-08 & 6.27e-06 & 4.90e-08 & $\cdots$ & 4.79e-05 & 18.631 & 0.149 & 0.052 & -0.029 & 2.98e+03 \\
33 & 12 & 71.7623 & -72.6490 & $\cdots$ & 72.8803 & -72.7133 & 0.125 & 51235 & 3.77e-05 & 4.31e-08 & 4.17e-06 & $\cdots$ & 1.05e-04 & 18.614 & 0.195 & -0.066 & -0.063 & 2.94e+03 \\
34 & 1 & 77.2652 & -74.0161 & $\cdots$ & 78.4914 & -74.0496 & 0.125 & 37072 & 7.12e-07 & 1.49e-07 & 2.63e-06 & $\cdots$ & 7.43e-05 & 18.628 & 0.205 & -0.073 & -0.079 & 2.60e+03 \\
$\vdots$ &$\vdots$ &$\vdots$  \\
\hline
\end{tabular}
\\
Table notes:
(1) J2000 coordinates for the 4 corners of subregions. 
(2) Effective area used for the SFH analysis. 
(3) Number of stars in the PSF catalog. 
(4) \SFRt\ for the 16 age bins. 
(5) True distance moduli and $V$-band extinctions derived from the color-magnitude shifts.
(6) Metallicity shifts at the two extremes of the age interval.
(7) Likelihood.
\end{scriptsize}
\end{table*}


\end{document}